\documentclass[12pt,prb,superscriptaddress,preprintnumbers,amsmath,amssymb]{revtex4-2}

\usepackage{latexsym}
\usepackage{amssymb}
\usepackage{amsmath}
\usepackage{bm}
\usepackage{amsfonts}
\usepackage{bbold}
\usepackage{slashed}
\usepackage{appendix}

\usepackage{graphicx} 
\usepackage{dcolumn} 

\newcommand{\eps}{\epsilon}

\newcommand{\be}{\mathbf{e}}
\newcommand{\bk}{\mathbf{k}}

\newcommand{\bq}{\mathbf{q}}
\newcommand{\bQ}{\mathbf{Q}}
\newcommand{\bR}{\mathbf{R}}

\newcommand{\bzero}{\mathbf{0}}

\newcommand{\cO}{{\cal O}}

\newcommand{\mbk}{{-\mathbf{k}}}
\newcommand{\bkq}{{\mathbf{k}+\mathbf{q}}}

\newcommand{\ellnot}{{-\ell}}

\newcommand{\gmk}{g^-_\mathbf{k}}
\newcommand{\hmk}{h^-_\mathbf{k}}
\newcommand{\hpk}{h^+_\mathbf{k}}
\newcommand{\emk}{e^-_\mathbf{k}}
\newcommand{\epk}{e^+_\mathbf{k}}

\newcommand{\chit}{\tilde{\chi}}

\newcommand{\up}{\uparrow}
\newcommand{\down}{\downarrow}
\newcommand{\tr}{\rm tr}

\begin{document}

\title{Spin stiffness, spectral weight, and Landau damping of magnons \\
in metallic spiral magnets}

\author{Pietro M.~Bonetti}
\affiliation{Max Planck Institute for Solid State Research,
 D-70569 Stuttgart, Germany}
\author{Walter Metzner}
\affiliation{Max Planck Institute for Solid State Research,
 D-70569 Stuttgart, Germany}

\date{\today}

\begin{abstract}
We analyze the properties of magnons in metallic electron systems with spiral magnetic order. Our analysis is based on the random phase approximation for the susceptibilities of tight binding electrons with a local Hubbard interaction in two or three dimensions. We identify three magnon branches from poles in the susceptibilities, one associated with in-plane, the other two associated with out-of-plane fluctuations of the spiral order parameter. We derive general expressions for the spin stiffnesses and the spectral weights of the magnon modes, from which also the magnon velocities can be obtained. Moreover, we determine the size of the decay rates of the magnons due to Landau damping. While the decay rate of the in-plane mode is of the order of its excitation energy, the decay rate of the out-of-plane mode is smaller so that these modes are asymptotically stable excitations even in the presence of Landau damping. 
\end{abstract}

\maketitle


\section{Introduction}

The Goldstone theorem predicts the emergence of gapless collective modes whenever a continuous symmetry of a physical system is spontaneously broken \cite{goldstone61}.
These Goldstone modes are ubiquitous in nature, and play a prominent role in particle physics and condensed matter physics alike.
In solids, the most important examples for Goldstone modes are phonons associated with the broken translation invariance in a crystal, and magnons in a magnetic state with broken SU(2) spin rotation invariance \cite{anderson}.

Usually Goldstone modes are asymptotically stable quasi-particles, that is, their decay rate (or ``damping'') is much smaller than their excitation energy, at least in the low energy limit. For example, a hydrodynamic theory of Goldstone modes in ferromagnets and antiferromagnets suggests that the decay rate of magnons is proportional to the square of the excitation energy \cite{halperin69}. The same behavior was found for a non-collinear helical spin arrangement \cite{halperin77}. The hydrodynamic theory for the decay rate has been confirmed by microscopic calculations for magnetic insulators such as the Heisenberg model \cite{harris70}.
In {\em metallic}\/ systems an additional low-energy decay channel exists via the excitation of particle-hole pairs near the Fermi surface, the socalled {\em Landau damping}\/ \cite{landau46}. In a N\'eel antiferromagnet this mechanism can lead to an enhanced decay rate of magnons proportional to their excitation energy \cite{sachdev95}.

In this paper we analyze the Goldstone modes, that is, magnons in a metallic system with planar spiral magnetic oder. Spiral magnetic states have been obtained in extended parameter regimes of the two-dimensional Hubbard and $t-J$ models at low and moderate hole doping away from half-filling, and they have been discussed as candidates for the incommensurate magnetic states observed in cuprate high temperature superconductors \cite{shraiman89, machida89, dombre90, fresard91, chubukov92, chubukov95, kotov04, igoshev10, yamase16, eberlein16, mitscherling18, bonetti20}. They compete with charge-spin stripe order \cite{qin21} and may coexist with $d$-wave superconductivity \cite{yamase16,sushkov04}.
Spiral magnetic order has also been observed in three dimensional correlated electron systems, for example in $\rm La_{1-x} X_x Mn O_3$ with X = Ba, Ca, Sr, etc.\ \cite{inoue95}, in $\rm V_{2-y} O_3$ \cite{bao93}, and in $\rm SrFeO_3$ \cite{takeda72, ishiwata20}.

A spiral state breaks the SU(2) spin rotation invariance completely so that no residual continuous symmetry survives. As a consequence, three distinct magnon branches emerge -- one more than for ferromagnetic or N\'eel-type antiferromagnetic states which remain symmetric under rotations around an axis parallel to the spin orientation.
One mode corresponds to spin fluctuations within the plane defined by the spiral order, while the other two correspond to out-of-plane fluctuations. The energy-momentum dispersion of all three magnon branches is linear \cite{rastelli85, chandra90, shraiman92, kampf96}.

We compute the spin susceptibilities in the spiral state in a random phase approximation (RPA) applied to itinerant electrons with a Hubbard interaction. In combination with a mean-field calculation of the order parameter, the RPA is a conserving approximation in the sense of Baym and Kadanoff \cite{baym61}, which is expected to capture the structure of collective modes without artifacts. Expanding the inverse in-plane and out-of-plane susceptibilities for small frequencies and momenta near the Goldstone points, we derive expressions for the spin stiffness, the spectral weight, and the damping of the magnons.
The Landau damping of the in-plane mode has the same momentum and frequency dependence as in a N\'eel state. However, the Landau damping of the out-of-plane modes is much smaller than in a N\'eel state, so that these modes are well defined, asymptotically stable quasi-particles.

The paper is structured as follows. In Sec.~II we summarize basic properties of the spiral state and we introduce a convenient rotated spin reference frame. Sec.~III contains a comprehensive analysis of the Goldstone mode poles in the RPA susceptibilities. The general analysis is complemented by a numerical evaluation for the two-dimensional Hubbard model. A conclusion in Sec.~IV closes the presentation.


\section{Spiral state}

A planar spiral antiferromagnetic state oriented in the $xy$-plane is characterized by an average magnetization of the form
\begin{equation} \label{eq:spiralmag}
 \langle \mathbf{S}_j \rangle =
 m \left[ \cos\left(\bQ \cdot \bR_j\right) \be_1 + 
 \sin\left(\bQ \cdot \bR_j\right) \be_2 \right] \, ,
\end{equation}
where $m$ is the magnetization amplitude, $\bR_j$ is the real space position of the lattice site $j$, and $\be_{\alpha}$ is a unit vector in the $\alpha$-direction, with $\alpha = 1,2,3$ corresponding to $x,y,z$, respectively. For SU(2) symmetric systems the $xy$ orientation of the magnetization is degenerate with an orientation along any other plane. $\bQ$ is a fixed wave vector.
Our general analytic results on the Goldstone modes are valid both in two and three spatial dimensions. 
Some numerical results are presented specifically for two-dimensional systems with ordering wave vectors of the form $\bQ = (\pi-2\pi\eta,\pi)$.

In an itinerant electron system the three components of the spin operator are given by
\begin{equation} \label{eq:spinitinerant}
 S^\alpha_j = \frac{1}{2} \sum_{s,s'=\up,\down} 
 c^\dagger_{j,s}\,\sigma^\alpha_{ss'}\,c_{j,s'} \, ,
\end{equation}
where $\sigma^\alpha$ with $\alpha = 1,2,3$ are the Pauli matrices, and $c_{j,s}^\dagger$ ($c_{j,s}$) are  electron creation (annihilation) operators at site $j$ with spin projection $s$. In momentum space, spiral order as in Eq.~(\ref{eq:spiralmag}) corresponds to anomalous expectation values
$\langle a^\dagger_{\bk,\up} a_{\bk+\bQ,\down}\big\rangle$, 
where $a_{\bk,s}^\dagger$ ($a_{\bk,s}$) creates (annihilates) electrons with momentum $\bk$ and spin orientation $s$. The momentum integral
\begin{equation}
 \int_\bk \,
 \langle a^\dagger_{\bk,\up} a_{\bk+\bQ,\down}\big\rangle = m
\end{equation}
determines the magnetization amplitude in Eq.~(\ref{eq:spiralmag}).
Here and in the following we use the short-hand notation
$\int_\bk = \int \frac{d^d\bk}{(2\pi)^d}$ for $d$-dimensional momentum integrals.

It is convenient to use a locally rotated spin reference frame \cite{kampf96}, corresponding to rotated fermion operators
\begin{equation} \label{eq:rotframe}
 \tilde c_j =
 e^{-\frac{i}{2}\bQ\cdot\bR_j} e^{\frac{i}{2}\bQ\cdot\bR_j \sigma^3} c_j \, , \quad
 \tilde c_j^\dagger =
 c_j^\dagger \, e^{-\frac{i}{2}\bQ\cdot\bR_j \sigma^3} e^{\frac{i}{2}\bQ\cdot\bR_j} \, ,
\end{equation}
where $c_j = (c_{j,\up},c_{j,\down})$ and $\tilde c_j = (\tilde c_{j,\up},\tilde c_{j,\down})$ are spinors with spin up and spin down components.
In this basis, the spiral state assumes the form of ferromagnetic order, with all the spins pointing along the $\be_1$ axis:
\begin{equation}
 \langle\tilde{S}^\alpha_j\rangle =
 \frac{1}{2} \big\langle \tilde c^\dagger_j \sigma^\alpha \tilde c_j \big\rangle =
 m \delta_{\alpha,1} \, .
\end{equation}
In momentum representation, the spin dependent phase factors in Eq.~(\ref{eq:rotframe}) correspond to momentum shifts, such that the Fourier transform of $\tilde c_j$ has the form
$\tilde a_\bk = (\tilde a_{\bk,\up},\tilde a_{\bk,\down}) = (a_{\bk,\up},a_{\bk+\bQ,\down})$.

In the rotated spinor basis, the mean-field Matsubara Green's function has the simple matrix form
\begin{equation} \label{eq:spiralGf}
 \tilde G(\bk,\nu) = 
  \left( \begin{array}{cc}
  i\nu-\xi_{\bk} & - \Delta \\ - \Delta & i\nu-\xi_{\bk+\bQ}
  \end{array} \right)^{-1} \, ,
\end{equation}
where $\xi_\bk = \eps_\bk - \mu$ with the single-particle dispersion $\eps_\bk$ and the chemical potential $\mu$, while $\Delta$ is the magnetic gap associated with the spiral order.
Diagonalizing the matrix in Eq.~(\ref{eq:spiralGf}), one obtains the quasi-particle energies
\begin{equation}
 E_{\bk}^\pm = g_{\bk} \pm \sqrt{h_{\bk}^2 + \Delta^2} \, ,
\end{equation}
where $g_\bk = \frac{1}{2}(\xi_\bk + \xi_{\bk+\bQ})$ and
$h_\bk = \frac{1}{2}(\xi_\bk - \xi_{\bk+\bQ})$.
The Green's function can be written as a linear combination of the quasi-particle poles,
\begin{equation} \label{eq:spiralGfcomf}
 \tilde G(\bk,\nu) = \frac{1}{2} \sum_{\ell=\pm} 
  \, \frac{u^\ell_{\bk}}{i\nu-E^\ell_{\bk}} \, ,
\end{equation}
with the coefficients
\begin{equation} \label{eq:ukell}
 u_\bk^\ell = \sigma^0 + \ell \, \frac{h_\bk}{e_\bk} \sigma^3 +
 \ell \, \frac{\Delta}{e_\bk} \sigma^1 \, , 
\end{equation}
where $\sigma^0$ is the $2\times2$ unit matrix and $e_\bk = \sqrt{h_\bk^2 + \Delta^2}$.

Within mean-field theory applied to the Hubbard model with a repulsive Hubbard interaction $U$, the magnetic gap is determined self-consistently by the gap equation
\begin{equation} \label{eq:gapequation}
 \Delta = - U \int_\bk T\sum_\nu \tilde G_{\up\down}(\bk,\nu) =
  U \int_\bk \frac{\Delta}{2e_\bk} \left[f(E^-_\bk)-f(E^+_\bk)\right] \, ,
\end{equation}
where $f(x)=(e^{x/T}+1)^{-1}$ is the Fermi function, and it is related to the magnetization amplitude by the simple relation $\Delta = Um$.

In a spiral state with a generic wave vector $\bQ$, spin and charge susceptibilities are coupled already on RPA level \cite{kampf96}. It is convenient to combine spin and charge variables by defining the charge-spin operator
\begin{equation}
 S_j^a = \frac{1}{2} \sum_{s,s'=\up,\down}
 c^\dagger_{j,s}\,\sigma^a_{ss'}\,c_{j,s'} \, ,
\end{equation}
with $a \in \{0,1,2,3\}$, where $\sigma^0$ is the unit matrix and $\sigma^\alpha$ with $\alpha \in \{1,2,3\}$ are the Pauli matrices. To treat spin and charge with the same conventions, the operator $S_j^0$ is defined as one half of the usual charge operator.
We define a combined imaginary-time charge-spin susceptibility
$\chi_{jj'}^{ab}(\tau) = \langle {\cal T} S_j^a(\tau) S_{j'}^b(0) \rangle$, where $\tau$ is the time variable and $\cal T$ the time-ordering operator.
Fourier transforming from imaginary time to imaginary (Matsubara) frequency representation, and continuing analytically to the real frequency axis, $i\Omega \to \omega + i0^+$, one obtains the retarded susceptibility which we denote as $\chi^{ab}_{jj'}(\omega)$.

The spin and charge susceptibilities can be computed within the rotated reference frame and then rotated back to the physical basis as \cite{kampf96}
\begin{equation} \label{eq:rotchireal}
 \chi^{ab}_{jj'}(\omega) =
 \sum_{c,d}\left(R^\bQ_j\right)_{ac} \chit^{cd}_{jj'}(\omega) \,
 \left(R^\bQ_{j'}\right)_{db}^\dagger \, ,
\end{equation}
where $\chit^{cd}_{jj'}(\omega)$ is the susceptibility in the rotated basis. The rotation matrix $R^\bQ_j$ is given by
\begin{equation}
 R^\bQ_j = \left(
 \begin{array}{cccc}
  1 & 0 & 0 & 0 \\
  0 & \cos\left(\bQ\cdot\bR_j\right) & -\sin\left(\bQ\cdot\bR_j\right) & 0 \\
  0 & \sin\left(\bQ\cdot\bR_j\right) & \cos\left(\bQ\cdot\bR_j\right) & 0 \\
  0 & 0 & 0 & 1
 \end{array} \right) \, .
\end{equation}
While $\chit^{cd}_{jj'}(\omega)$ is translation invariant, components of $\chi^{ab}_{jj'}(\omega)$ with $a,b \in \{1,2\}$ are generally not. Their momentum representation $\chi^{ab}(\bq,\bq',\omega)$ therefore involves not only momentum diagonal terms with $\bq' = \bq$, but also off-diagonal terms with $\bq' = \bq \pm \bQ$ (only for $a \neq b$) and $\bq' = \bq \pm 2\bQ$.
We denote the momentum-diagonal part of the susceptibilities $\chi^{ab}(\bq,\bq',\omega)$ by $\chi^{ab}(\bq,\omega)$.
Fourier transforming Eq.~(\ref{eq:rotchireal}), we obtain the following linear relations between $\chi^{aa}(\bq,\omega)$ and $\chit^{ab}(\bq,\omega)$,
\begin{eqnarray}
 \chi^{00}(\bq,\omega) &=& \chit^{00}(\bq,\omega) \\
 \chi^{11}(\bq,\omega) &=& \chi^{22}(\bq,\omega) \nonumber \\
 &=& \frac{1}{4} \big[
 \chit^{11}(\bq+\bQ,\omega) + \chit^{11}(\bq-\bQ,\omega) +
 \chit^{22}(\bq+\bQ,\omega) + \chit^{22}(\bq-\bQ,\omega) \nonumber \\
 && \quad + 2i \, \chit^{12}(\bq+\bQ,\omega) +
    2i \, \chit^{21}(\bq-\bQ,\omega) \big] \nonumber \\
 &=& \chit^{-+}(\bq+\bQ,\omega) + \chit^{+-}(\bq-\bQ,\omega) \, ,
 \label{eq:chi11phys} \\
 \chi^{33}(\bq,\omega) &=& \chit^{33}(\bq,\omega) \, ,
\end{eqnarray}
where we have used $\chit^{21}=-\chit^{12}$ (see Table~\ref{tab:symmetries} and Appendix~\ref{app:chi0symm}), and we have defined
\begin{equation}
    \chit^{+-}(\bq,\omega)=\langle \tilde{S}^+_{-\bq,-\omega}\tilde{S}^-_{\bq,\omega}\rangle,
\end{equation}
with $\tilde{S}^\pm=(\tilde{S}^1\pm i\tilde{S}^2)/2$. While these relations hold both for real and imaginary frequencies, we denote real frequency arguments by $\omega$ in the following sections.
For $a = b$, the only off-diagonal (in momentum) susceptibilities are
\begin{eqnarray}
 \chi^{11}(\bq,\bq \pm 2\bQ,\omega) &=& \frac{1}{4} \left[
 \chit^{11}(\bq \mp \bQ,\omega) - \chit^{22}(\bq \mp \bQ,\omega) \right] , \\
 \chi^{22}(\bq,\bq \pm 2\bQ,\omega) &=& \frac{1}{4} \left[
 \chit^{22}(\bq \mp \bQ,\omega) - \chit^{11}(\bq \mp \bQ,\omega) \right] .
\end{eqnarray}
In the special case of a N\'eel state there are no momentum off-diagonal susceptibilities and the above relations for $\chi^{11}(\bq,\bq',\omega)$ and $\chi^{22}(\bq,\bq',\omega)$ are not valid. We will discuss the N\'eel case separately in Sec.~\ref{sec:Neel}.


\section{Susceptibilities and Goldstone modes} \label{sec:RPAsusc}

Within the RPA, the charge-spin susceptibility of the Hubbard model in the rotated basis is given by
\begin{equation} \label{eq:chitRPA}
 \chit(q) = \chit_0 (q) \left[ \mathbb{1} - \Gamma_0\chit_0(q) \right]^{-1} \, ,
\end{equation}
where $\mathbb{1}$ is the four-dimensional unit matrix,
$\Gamma_0 = 2 \, \mathrm{diag}(-U,U,U,U)$, and the bare susceptibility components on the real frequency axis can be expressed as \cite{negele87}
\begin{equation}
 \chit^{ab}_0(\bq,\omega) = - \frac{1}{4} \int_\bk T \sum_\nu \tr \big[ 
 \sigma^a \, {\tilde G}(\bk+\bq,\nu+\Omega) \,
 \sigma^b \, {\tilde G}(\bk,\nu) \big] \Big\rvert_{i\Omega\to\omega+i0^+}.
\end{equation}
Using Eq.~\eqref{eq:spiralGfcomf}, one can easily perform the Matsubara sum to obtain
\begin{equation} \label{eq:chit0}
 \chit^{ab}_0(\bq,\omega) =
 - \frac{1}{8}\int_\bk \sum_{\ell,\ell'} A^{ab}_{\ell\ell'}(\bk,\bq)
 F_{\ell\ell'}(\bk,\bq),
\end{equation}
with 
\begin{equation} \label{eq:F}
 F_{\ell\ell'}(\bk,\bq,\omega) =
 \frac{f(E^\ell_\bk) - f(E^{\ell'}_\bkq)}{\omega + i0^+ + E^\ell_\bk - E^{\ell'}_\bkq},
\end{equation}
and the coherence factors $A^{ab}_{\ell\ell'}(\bk,\bq)$ defined as
\begin{equation} \label{eq:Acalc}
 A^{ab}_{\ell\ell'}(\bk,\bq) = \frac{1}{2}
 \tr\left[\sigma^a\,u_\bk^\ell\,\sigma^b\,u_{\bk+\bq}^{\ell'}\right] \, ,
\end{equation}
with $u_\bk^\ell$ from Eq.~\eqref{eq:ukell}. Explicit expressions for the coherence factors are listed in Appendix~\ref{app:cohfac}.

The coherence factors are either purely real or purely imaginary, depending on $a$ and $b$. The functions $F_{\ell\ell'}(\bk,\bq,\omega)$ have a real part and an imaginary part proportional to a $\delta$-function. To distinguish the corresponding contributions to $\chit^{ab}_0(\bq,\omega)$, we refer to the contribution coming from the real part of $F_{\ell\ell'}(\bk,\bq,\omega)$ as $\chit^{ab}_{0r}(\bq,\omega)$, and the contribution from the imaginary part of $F_{\ell\ell'}(\bk,\bq,\omega)$ as $\chit^{ab}_{0i}(\bq,\omega)$. Note that $\chit^{ab}_{0r}(\bq,\omega)$ is imaginary and $\chit^{ab}_{0i}(\bq,\omega)$ is real if the corresponding coherence factor is imaginary. 

Before proceeding, we first note some symmetries of the bare susceptibilities.


\subsection{Symmetries of the bare susceptibilities}

The contributions $\chit_{0r}^{ab}$ and $\chit_{0i}^{ab}$ to $\chit_0^{ab}$ have a well defined parity under $\bq \to -\bq$. In Appendix~\ref{app:chi0symm} we show that the diagonal components of $\chit_{0r}^{ab}$ and the off-diagonal ones which do not involve either the 2- or the 3-component of the spin are symmetric, while the other off-diagonal elements are antisymmetric.
The sign change of $\chit_{0i}^{ab}(q)$ under $\bq \to -\bq$ is the opposite, that is, $\chit_{0i}^{ab}(q)$ is antisymmetric if $\chit_{0r}^{ab}(q)$ is symmetric and vice versa.
For a spiral wavevector $\bQ$ of the form $(\pi-2\pi\eta,\pi)$ all the susceptibilities are symmetric under $q_y \to -q_y$. This implies that those susceptibilities which are antisymmetric for $\bq \to -\bq$ are identically zero for $q_y=0$, and vanish in the limit of N\'eel order ($\eta \to 0$). Similarly, for a diagonal spiral $\bQ = (\pi-2\pi\eta,\pi-2\pi\eta)$ all the susceptibilities are symmetric for $q_x \leftrightarrow q_y$ and those which are antisymmetric in $\bq$ vanish for $q_x = q_y$.

The contributions $\chit_{0r}^{ab}$ and $\chit_{0i}^{ab}$ to $\chit_0^{ab}$ are also either symmetric or antisymmetric under the transformation $\omega \to -\omega$.
In Appendix~\ref{app:chi0symm} we show that among the functions $\chit_{0r}^{ab}$ all the diagonal parts and the off-diagonal ones which do not involve the 3-component of the spin are symmetric in $\omega$. The off-diagonal terms involving the 3-component of the spin are antisymmetric. $\chit_{0i}^{ab}(q)$ is antisymmetric under $\omega \to -\omega$ if $\chit_{0r}^{ab}(q)$ is symmetric and vice versa.

In Table \ref{tab:symmetries} we show a summary of the generic (for arbitrary $\bQ$) symmetries of the bare susceptibilities. Susceptibilities with real (imaginary) coherence factors are symmetric (antisymmetric) under the exchange $a \leftrightarrow b$.
\begin{table}[ht!]
\centering
\begin{tabular}{|c|c|c|c|c|}
        \hline
        $a,b$ & 0 & 1 & 2 & 3  \\
        \hline
        0 & $+,+,+$ & $+,+,+$ & $-,+,-$ & $-,-,+$ \\
        \hline
        1 & $+,+,+$ & $+,+,+$ & $-,+,-$ & $-,-,+$ \\
        \hline
        2 & $-,+,-$ & $-,+,-$ & $+,+,+$ & $+,-,-$ \\
        \hline
        3 & $-,-,+$ & $-,-,+$ & $+,-,-$ & $+,+,+$ \\
        \hline
\end{tabular}
\caption{Symmetries of the bare susceptibilities. The first sign in each field represents the sign change of $\chit_{0r}^{ab}(q)$ under $\bq \to -\bq$. The second one represents the sign change of ${\chit_{0r}^{ab}}(q)$ under $\omega \to -\omega$. The sign changes of ${\chit_{0i}^{ab}}(q)$ under $\bq \to -\bq$ or $\omega \to -\omega$ are just the opposite. The third sign in each field is the sign change of $\chit_0^{ab}(q)$ under the exchange $a \leftrightarrow b$.}
\label{tab:symmetries}
\end{table}
%


\subsection{Location of Goldstone modes}

We now locate the Goldstone modes in the spiral state by identifying divergencies of the rotated susceptibilities $\chit(\bq,\omega)$. 


\subsubsection{In-plane mode}

At $(\bq,\omega)=(\bzero,0)$ all the off-diagonal bare susceptibilities involving a 2-component of the spin vanish: $\chit_0^{20}$ and $\chit_0^{21}$ vanish because they are odd in $\bq$ for $\omega=0$, while $\chit_{0r}^{23}$ vanishes because it is odd in $\omega$. Moreover, $\chit_{0i}^{23}$ vanishes at $(\bq,\omega)=(\bzero,0)$ because the intraband coherence factor $A_{\ell\ell}^{23}(\bk,\bq)$ vanishes for $\bq=\bzero$, and there are generally no interband contributions to $\chit_{0i}^{ab}$ at low frequencies.
The RPA expression for the 22-component of the rotated susceptibility therefore takes the simple form
\begin{equation} \label{eq:chi22(0)}
 \chit^{22}(\bzero,0) =
 \frac{\chit^{22}_0(\bzero,0)}{1 - 2 U\chit^{22}_0(\bzero,0)}.
\end{equation}
Eqs.~(\ref{eq:chit0}) and (\ref{eq:A22}) yield
\begin{equation}
    \chit^{22}_0(\bzero,0) = \int_\bk \frac{f(E^-_\bk)-f(E^+_\bk)}{4e_\bk}.  
\end{equation}
Note that the limit $\bq \to \bzero$, $\omega \to 0$ is unique here, since only interband contributions ($\ell' \neq \ell$) contribute.
The denominator of Eq.~\eqref{eq:chi22(0)} vanishes when the gap equation \eqref{eq:gapequation} is fulfilled. Hence, $\chit^{22}(\bzero,0)$ diverges. From Eq.~\eqref{eq:chi11phys} we see that this divergence entails divergencies in the translation invariant part of the physical susceptibilities $\chi^{11}(\bq,\omega)$ and $\chi^{22}(\bq,\omega)$ at $(\bq,\omega) = (\pm\bQ,0)$. These divergencies are associated with a massless (that is, gapless) Goldstone mode in the $xy$ plane \cite{chubukov95}, in which the average magnetization is aligned.
By contrast, $\chit^{11}(\bzero,0)$ is finite, corresponding to a massive amplitude mode. 


\subsubsection{Out-of-plane modes}

At zero frequency and finite $\bq$ all the off-diagonal elements involving the 3-component of the spin vanish. The contributions $\chit_{0r}^{3b}$ with $b \neq 3$ vanish as they are antisymmetric in $\omega$, while the contributions $\chit_{0i}^{ab}$ generally vanish for $\omega = 0$ at finite $\bq$. Hence, in this limit, the 33-component of the susceptibility also takes a simple form,
\begin{equation} \label{eq:chi33(q,0)}
 \chit^{33}(\bq,0) =
 \frac{\chit^{33}_0(\bq,0)}{1 - 2 U \chit^{33}_0(\bq,0)} \, .
\end{equation}
In Appendix \ref{app:chi33} we show that
\begin{equation} \label{eq:chi33_0(q,0)}
 \chit_0^{33}(\pm\bQ,0) = \int_\bk \frac{f(E^-_\bk)-f(E^+_\bk)}{4e_\bk} =
 \chit_0^{22}(\bzero,0) \, ,
\end{equation}
such that the denominator of Eq.~(\ref{eq:chi33(q,0)}) vanishes for $\bq = \pm \bQ$ if the gap equation is fulfilled.
Therefore, the 33-component of the susceptibility $\chi^{33} = \tilde\chi^{33}$ contains two Goldstone modes located at $\bq = \pm \bQ$, in this case associated with spin fluctuations out of the magnetization plane~\cite{chubukov95}. 


\subsection{Properties of Goldstone modes}

The susceptibilities containing a Goldstone mode pole can be expanded around the zero in the denominator as
\begin{equation}
 \chit^{aa}(\bq,\omega) \sim \frac{m^2}
 {J^{(a)}_{\alpha\beta}\,(q_\alpha-Q^{(a)}_{\alpha}) (q_\beta-Q^{(a)}_{\beta})
 - Z^{(a)} \,\omega^2 + \mbox{\it damping}} \, ,
\end{equation}
where $\bQ^{(a)} = (Q^{(a)}_x,Q^{(a)}_y)$ are the wave vectors of the Goldstone modes ($\bQ^{(2)} = \bzero$, $\bQ^{(3)} = \pm\bQ$), and $m$ is the magnetization amplitude defined in Eq.~\eqref{eq:spiralmag}.
The coefficients $J^{(a)}_{\alpha\beta}$ determine the diagonal and (if non-zero) off-diagonal components of the spin stiffness. The ratios $m^2/Z^{(a)}$ define the spectral weights of the Goldstone modes, and the ratios $J^{(a)}_{\alpha\beta}/Z^{(a)}$ their velocities. We refer to $Z^{(a)}$ as spectral weight factors or simply ``$Z$-factors''. 
The momentum and frequency dependence of the imaginary damping term will be specified below.

$Z^{(a)}$ can be extracted from the susceptibilities as
\begin{equation} \label{eq:Za}
 Z^{(a)} = - \frac{m^2}{2} \, \partial^2_\omega \left.
 \left({\rm Re} \, \frac{1}{\chit^{aa}(\bQ^{(a)},\omega)}\right) \right|_{\omega=0} \, ,
\end{equation}
where $\partial_\omega^2$ denotes the second derivative with respect to the frequency $\omega$. 
Similarly, the spin stiffness $J^{(a)}$ can be evaluated as
\begin{equation} \label{eq:Ja}
 J^{(a)}_{\alpha\beta} = \frac{m^2}{2} \, \partial^2_{q_\alpha q_\beta} \left.
 \left(\frac{1}{\chit^{aa}(\bq,0)}\right) \right|_{\bq=\bQ^{(a)}} \, ,
\end{equation}
with $\partial^2_{q_\alpha q_\beta} =
 \frac{\partial^2}{\partial q_\alpha \partial q_\beta}$.


\subsubsection{In-plane mode}

The off-diagonal susceptibilities connecting the sectors $0$ and $1$ to the sectors $2$ and $3$ vanish for $\bq = 0$. The contributions $\chit_{0r}^{ab}$ vanish due to their antisymmetry in $\bq$, while $\chit_{0i}^{ab}$ generally vanishes at $\bq=0$ and low finite frequency. To compute $Z^{(2)}$, it is thus sufficient to invert the $2\times2$ matrix involving only $\chit_0^{ab}$ with $a,b \in \{2,3\}$ to obtain $\chit^{22}(\bzero,\omega)$ from the RPA expression \eqref{eq:chitRPA}.
Expanding for small $\omega$, using
$\chit_0^{aa}(\bzero,\omega) = \chit_0^{aa}(\bzero,0) + {\cal O}(\omega^2)$ and
$\chi_0^{23}(\bzero,\omega) = - \chi_0^{32}(\bzero,\omega) = {\cal O}(\omega)$, one obtains
\begin{equation} \label{eq:Z2RPA}
 Z^{(2)} = 2\Delta^2 \left[
 \partial_\omega^2 \chit_0^{22}(\bzero,\omega) \big|_{\omega=0} 
 + \frac{4U}{1 - 2U\chit_0^{33}(\bzero,0)}
 \left|\partial_\omega\chit_0^{23}(\bzero,\omega)\big|_{\omega=0}\right|^2
 \right] \, .
\end{equation}

To compute the in-plane spin stiffness, one has to take into account only the components of the susceptibilities with $a,b \in \{0,1,2\}$, since the 3-component gets decoupled for $\omega = 0$. The function $\chit^{22}(\bq,0)$ can be extracted most efficiently from the RPA expression \eqref{eq:chitRPA} by using a suitable Schur complement for inverting the matrix (see Appendix \ref{app:chi22}). Expanding the matrix elements to second order in $\bq$, one obtains
\begin{equation} \label{eq:chit22inv}
 \frac{1}{\chit^{22}(\bq,0)} = 2U \bigg[
 1 - 2U \chit_0^{22}(\bq,0) - 2U \sum_{a,b=0,1}
 \chit_0^{2a}(\bq,0) \bar\Gamma^{ab}(\bzero,0) \, \chit_0^{b2}(\bq,0) \bigg]
 + {\cal O}(|\bq|^3) \, .
\end{equation}
The matrix $\bar\Gamma(q)$ represents the RPA effective interaction in the subspace spanned by the charge channel and the spin channel in $x$-direction,
\begin{equation}
 \bar\Gamma(q) =
 \left[ \mathbb{1}_2 -
 \left( \begin{array}{cc} -2U & 0 \\ 0 & 2U \end{array} \right)
 \left( \begin{array}{cc} \chit_0^{00}(q) & \chit_0^{01}(q) \\
 \chit_0^{10}(q) & \chit_0^{11}(q) \end{array} \right)
 \right]^{-1}
 \left( \begin{array}{cc} -2U & 0 \\ 0 & 2U \end{array} \right) \, .
\end{equation}
The matrix elements $\bar\Gamma^{ab}(q)$ with $a,b \in \{0,1\}$ are all finite for $q = (\bzero,0)$.

Inserting Eq.~\eqref{eq:chit22inv} into Eq.~\eqref{eq:Ja}, the stiffness of the in-plane Goldstone mode can then be expressed as
\begin{equation} \label{eq:J2RPA}
 J^{(2)}_{\alpha\beta} = - 2 \Delta^2 \Big[
 \partial^2_{q_\alpha q_\beta}\chit_0^{22}(\bq,0) \big|_{\bq=0} +
 2 \sum_{a,b=0,1} \left( \partial_{q_\alpha}\chit_0^{2a}(\bq,0)\big|_{\bq=0} \right)
 \bar\Gamma^{ab}(\bzero,0)
 \left( \partial_{q_\beta}\chit_0^{b2}(\bq,0)\big|_{\bq=0} \right) \Big] \, .
\end{equation}

For a two-dimensional spiral state with a wavevector of the form $\bQ = (\pi-2\pi\eta,\pi)$, the spin stiffness is diagonal in the spatial indices, that is, $J^{(2)}_{xy} = J^{(2)}_{yx}=0$, while $J^{(2)}_{xx} \neq J^{(2)}_{yy}$ for $\eta > 0$. The second term in Eq.~\eqref{eq:J2RPA} is nonzero only for $\alpha=\beta=y$.
By contrast, for a diagonal spiral with $\bQ=(\pi-2\pi\eta,\pi-2\pi\eta)$, we have $J^{(2)}_{xy} = J^{(2)}_{yx} \neq 0$, and $J^{(2)}_{xx} = J^{(2)}_{yy}$. In this case the second term in Eq.~\eqref{eq:J2RPA} does not depend on $\alpha$ and $\beta$.

We now determine the momentum and frequency dependence of the leading imaginary term describing the damping of the in-plane Goldstone mode for small $\bq$. Imaginary contributions to the diagonal susceptibilities arise from the $\delta$-function contributions $\chit_{0i}^{ab}$ to $\chit_0^{ab}$. For small frequencies (and small $\bq$), only intraband terms ($\ell=\ell'$) contribute since $E_\bk^+ - E_\bk^- > 2\Delta$.
We expand the imaginary part of $1/\chit^{22}(\bq,\omega)$ for small $\bq$, keeping the ratio $\hat\omega = \omega/|\bq|$ fixed.
The coupling to the 3-component can be neglected, since the intraband coherence factor $A_{\ell\ell}^{23}(\bk-\bq/2,\bq)$ is of order $|\bq|^2$ for small $\bq$.
Hence, for the imaginary part, the expansion Eq.~\eqref{eq:chit22inv} can be generalized to
\begin{equation} \label{eq:chit22ominv}
 {\rm Im} \, \frac{1}{\chit^{22}(\bq,\omega)} = - 4U^2 \bigg[
 \chit_{0i}^{22}(\bq,\omega) + {\rm Im} \! \sum_{a,b=0,1}
 \chit_0^{2a}(\bq,\omega) \bar\Gamma^{ab}(\bzero,0) \, \chit_0^{b2}(\bq,\omega)
 \bigg] + {\cal O}(|\bq|^3)
\end{equation}
for small $\bq$ and fixed finite $\hat\omega$. We will now show that both terms in Eq.~\eqref{eq:chit22ominv} are of order $|\bq|^2$ at fixed $\hat\omega$.

Shifting the integration variable $\bk$ in Eq.~\eqref{eq:chit0} by $-\bq/2$, the imaginary part of $\chit_0^{22}(\bq,\omega)$ can be written as
\begin{equation} \label{eq:chit0i22}
 \chit_{0i}^{22}(\bq,\omega) = \frac{i\pi}{8} \int_\bk \sum_{\ell,\ell'}
 A_{\ell\ell'}^{22}(\bk-\bq/2,\bq) \,
 \big[ f(E_{\bk-\bq/2}^\ell) - f(E_{\bk+\bq/2}^{\ell'}) \big] \,
 \delta(\omega + E_{\bk-\bq/2}^\ell - E_{\bk+\bq/2}^{\ell'}) \, .
\end{equation}
For small frequencies, only intraband terms contribute. The intraband coherence factor
\begin{equation}
 A_{\ell\ell}^{22}(\bk-\bq/2,\bq) = 
 1 - \frac{h_{\bk-\bq/2} h_{\bk+\bq/2} + \Delta^2}{e_{\bk-\bq/2} e_{\bk+\bq/2}}
\end{equation}
is of order $|\bq|^2$ for small $\bq$. Expanding
$E_{\bk+\bq/2}^\ell - E_{\bk-\bq/2}^\ell =
\bq \cdot \nabla_\bk E_\bk^\ell + {\cO}(|\bq|^3)$,
and using $\delta(|\bq|x) = |\bq|^{-1} \delta(x)$,
we find that $\chit_{0i}^{22}(\bq,\omega)$ is of order $|\bq|^2$.

Since $\bar\Gamma^{ab}(\bzero,0)$ is real, the second term in Eq.~\eqref{eq:chit22ominv} receives contributions from the cross terms
$\chit_{0r}^{2a}(\bq,\omega) \bar\Gamma^{ab}(\bzero,0) \, \chit_{0i}^{b2}(\bq,\omega)$ and
$\chit_{0i}^{2a}(\bq,\omega) \bar\Gamma^{ab}(\bzero,0) \, \chit_{0r}^{b2}(\bq,\omega)$.
For small $\omega$, only intraband terms contribute to $\chit_{0i}^{2a}(\bq,\omega)$ and $\chit_{0i}^{b2}(\bq,\omega)$. Both are of order $\bq$ for small $\bq$ at fixed $\hat\omega$, because the intraband coherence factors $A_{\ell\ell}^{02}(\bk,\bq) = - A_{\ell\ell}^{20}(\bk,\bq)$ and $A_{\ell\ell}^{12}(\bk,\bq) = - A_{\ell\ell}^{21}(\bk,\bq)$ are of order $\bq$. Moreover, $\chit_{0r}^{2a}(\bq,\omega)$ and $\chit_{0r}^{b2}(\bq,\omega)$ are antisymmetric in $\bq$ and thus of order $\bq$, too. Hence, the second term in Eq.~\eqref{eq:chit22ominv} is of order $|\bq|^2$.

In summary, we have shown that the damping term of the in-plane Goldstone mode has the scaling form
\begin{equation} \label{eq:damping2}
 {\rm Im} \, \frac{m^2}{\chit^{22}(\bq,\omega)} =
 - |\bq|^2 \gamma(\hat\bq,\hat\omega) + \cO(|\bq|^3) \, ,
\end{equation}
where $\gamma(\hat\bq,\hat\omega)$ is a function of $\hat\bq = \bq/|\bq|$ and $\hat\omega = \omega/|\bq|$. The scaling function $\gamma(\hat\bq,\hat\omega)$ has the same sign as $\hat\omega$, and it vanishes for $\hat\omega = 0$.
The damping of the in-plane mode thus has the same form as the Landau damping of the two Goldstone modes in a N\'eel antiferromagnet \cite{sachdev95}.
It is of the same order as the leading real terms near the Goldstone pole. Hence, the damping of the in-plane Goldstone mode is of the same order as its excitation energy, that is, of order $|\bq|$. Asymptotically stable low-energy quasi-particles require damping rates that vanish faster than their excitation energy in the low-energy limit. The in-plane Goldstone mode in a metallic spiral state and the Goldstone mode in a metallic N\'eel state violate this criterion, albeit only marginally.


\subsubsection{Out-of-plane mode}

We derive the RPA expression for the out-of-plane spectral weight factor $Z^{(3)}$ in a rotated spin basis spanned by $S_j^\pm = \frac{1}{2} (S_j^1 \pm i S_j^2)$ instead of $S_j^1$ and $S_j^2$. 
The coherence factors in this basis, $A_{\ell\ell'}^{ab}$ with $a,b \in \{ 0,+,-,3 \}$, are all real.
The matrix elements of the bare interaction matrix $\Gamma_0$ with indices $+$ and $-$ are $\Gamma_0^{+-} = \Gamma_0^{-+} = 4U$ and $\Gamma_0^{++} = \Gamma_0^{--} = 0$. 
The components $\chi^{+-}$ and $\chi^{-+}$ of the physical susceptibility are diagonal in momentum space, while $\chi^{++}$ and $\chi^{--}$ are off-diagonal, with a momentum shift~$\bQ$.

For $\bq = \bQ$ and finite $\omega$ the 3-component of the spin couples to all the other spin components and the charge channel. The function $\chit^{33}(\bQ,\omega)$ can again be extracted from the RPA expression \eqref{eq:chitRPA} by using a suitable Schur complement. Expanding the matrix elements to second order in $\omega$, one obtains
\begin{equation} \label{eq:chit33inv}
 \frac{1}{\chit^{33}(\bQ,\omega)} = 2U \bigg[
 1 - 2U \chit_0^{33}(\bQ,\omega) - 2U \!\!\! \sum_{a,b=0,+,-}
 \chit_0^{3a}(\bQ,\omega) \bar\Gamma^{{\bar a}b}(\bQ,0) \, 
 \chit_0^{{\bar b}3}(\bQ,\omega) \bigg]
 + {\cal O}(\omega^3) \, ,
\end{equation}
where the bar over the indices $a$ and $b$ leaves the index $0$ unchanged, while it exchanges the indices $+$ and $-$. Here, the matrix $\bar\Gamma(q)$ represents the RPA effective interaction in the subspace spanned by the charge channel and the in-plane spin channel in the basis spanned by $S^+$ and $S^-$,
\begin{equation}
 \bar\Gamma(q) =
 \left[ \mathbb{1}_3 -
 \left( \begin{array}{ccc} -2U & 0 & 0 \\ 0 & 4U & 0 \\ 0 & 0 & 4U \end{array} \right)
 \left( \begin{array}{ccc} \chit_0^{00}(q) & \chit_0^{0-}(q) & \chit_0^{0+}(q) \\
 \chit_0^{+0}(q) & \chit_0^{+-}(q) & \chit_0^{++}(q) \\ 
 \chit_0^{-0}(q) & \chit_0^{--}(q) & \chit_0^{-+}(q) \end{array} \right)
 \right]^{-1}
 \left( \begin{array}{ccc} -2U & 0 & 0 \\ 0 & 4U & 0 \\ 0 & 0 & 4U \end{array} \right)
 \, .
\end{equation}
Inserting Eq.~\eqref{eq:chit33inv} into Eq.~\eqref{eq:Za}, the out-of-plane spectral weight factor can be expressed in the form
\begin{equation} \label{eq:Z3RPA}
 Z^{(3)} = 2\Delta^2 \Big[
 \partial_\omega^2\chit_0^{33}(\bQ,\omega)\big|_{\omega=0} +
 2 \sum_{a,b=0,+,-} \left(
 \partial_\omega \chit_0^{3a}(\bQ,\omega)\big|_{\omega=0} \right)
 \bar\Gamma^{\bar{a}b}(\bQ,0)
 \left( \partial_\omega\chit_0^{\bar{b}3}(\bQ,\omega)\big|_{\omega=0} \right)
 \Big] \, .
\end{equation}
This expression is real, because $\chit_{0i}^{33}(\bq,\omega)$ is antisymmetric in $\omega$, while $\chit_{0i}^{3a}(\bq,\omega)$ with $a \neq 3$ is symmetric.

The expression for the out-of-plane spin stiffness is comparatively simple, since all off-diagonal susceptibilities involving the 3-component of the spin vanish at $\omega = 0$. Expanding around $\bq = \pm \bQ$ one obtains
\begin{equation} \label{eq:J3RPA}
 J^{(3)}_{\alpha\beta} = - 2\Delta^2 \,
 \partial^2_{q_{\alpha\beta}}\chit_0^{33}(\bq,0)\big|_{\bq=\pm\bQ} \, .
\end{equation}
For the most common spiral states in two dimensions with wave vectors of the form $\bQ = (\pi,\pi-2\pi\eta)$ and $\bQ = (\pi-2\pi\eta,\pi-2\pi\eta)$, the spatial structure of $J^{(3)}_{\alpha\beta}$ is the same as for the in-plane stiffness $J^{(2)}_{\alpha\beta}$ discussed above.

We finally determine the asymptotic momentum and frequency dependence of the imaginary part of $1/\chit^{33}(\bq,\omega)$ for $\bq$ near $\pm\bQ$, which determines the damping of the out-of-plane Goldstone modes. We discuss the case $\bq \sim \bQ$. The behavior for $\bq \sim - \bQ$ is equivalent.

We first analyze the low frequency asymptotics for $\bq = \bQ$ and show that all contributions to the imaginary part of $1/\chit^{33}(\bQ,\omega)$ in Eq.~\eqref{eq:chit33inv} are of order $\omega^3$.
The first contribution is determined by the imaginary part of the bare out-of-plane spin susceptibility,
\begin{equation} \label{eq:chit0i33}
 \chit_{0i}^{33}(\bQ,\omega) = \frac{i\pi}{8} \int_\bk \sum_{\ell,\ell'}
 A_{\ell\ell'}^{33}(\bk,\bQ)
 \big[ f(E_\bk^\ell) - f(E_{\bk+\bQ}^{\ell'}) \big]
 \delta(\omega + E_\bk^\ell - E_{\bk+\bQ}^{\ell'}) \, .
\end{equation}
For small frequencies $\omega$, only momenta corresponding to small energies $E_\bk^\ell$ and $E_{\bk+\bQ}^{\ell'}$ of order $\omega$ contribute to the $\bk$-integral. These momenta are restricted to a small neighborhood of {\em hot spots}\/ $\bk_H$ defined by the equations
\begin{equation} \label{eq:hotspots1}
 E_{\bk_H}^\ell = E_{\bk_H+\bQ}^{\ell'} = 0 \, .
\end{equation}
Geometrically, the hot spots are the intersection points of the Fermi surface of $E_\bk^\ell$ and the $\bQ$-shifted Fermi surface of $E_\bk^{\ell'}$.
In our two-dimensional case studies (see below) we have only found intraband ($\ell=\ell'$) hot spots. While we cannot exclude the existence of interband hot spots in general, we restrict the subsequent analysis to intraband contributions.

For $\ell=\ell'$, the equations \eqref{eq:hotspots1} are equivalent to
\begin{equation} \label{eq:hotspots2}
 E_{\bk_H}^\ell = 0 \quad \mbox{and} \quad \xi_{\bk_H} = \xi_{\bk_H+2\bQ} \, .
\end{equation}
We note that for a N\'eel state, where $2\bQ$ is a reciprocal lattice vector, the second equation is always satisfied, so that all momenta on the Fermi surface of $E_{\bk}^\ell$ are hot spots.
The condition $\xi_{\bk_H} = \xi_{\bk_H+2\bQ}$ implies that $h_{\bk_H+\bQ} = - h_{\bk_H}$. As a direct consequence, we find that $A_{\ell\ell}^{33}(\bk_H,\bQ) = 0$ and also
$\nabla_\bk A_{\ell\ell}^{33}(\bk,\bQ) \big|_{\bk=\bk_H} = 0$.
Hence, the coherence factor leads to a strong suppression of $\chit_{0i}^{33}(\bQ,\omega)$ at low frequencies. For small $\omega$, the momenta $\bk$ contributing to the integral in Eq.~\eqref{eq:chit0i33} are situated at a distance of order $\omega$ away from the hot spots. For such momenta the coherence factor $A_{\ell\ell}^{33}(\bk,\bQ)$ is of order $\omega^2$, since $A_{\ell\ell}^{33}(\bk,\bQ)$ and also its gradient vanish at $\bk=\bk_H$.
Multiplying this with the usual factor $\omega$ coming from the difference of Fermi functions, we obtain
\begin{equation}
\chit_{0i}^{33}(\bQ,\omega) \propto \omega^3
\end{equation}
for small $\omega$.

We now turn to the second contribution to the imaginary part of $1/\chit^{33}(\bQ,\omega)$ in Eq.~\eqref{eq:chit33inv}, which involves the off-diagonal bare susceptibilities $\chit_0^{3a}$ and $\chit_0^{a3}$ with $a \in \{0,+,-\}$. Since the static RPA effective interaction in Eq.~\eqref{eq:chit33inv} is real, contributions to the imaginary part of $1/\chit^{33}(\bQ,\omega)$ are due to products of real and imaginary parts of the off-diagonal bare susceptibilities.
The real parts $\chit_{0r}^{3a}(\bQ,\omega)$ are antisymmetric in the frequency argument and thus of order $\omega$ for small $\omega$.
The coherence factors $A_{\ell\ell}^{3a}(\bk,\bQ)$ vanish at the hot spots, but their gradients $\nabla_\bk A_{\ell\ell}^{3a}(\bk,\bQ)$ are finite at $\bk=\bk_H$.
Hence, following the above arguments used to determine the low frequency dependence of $\chit_{0i}^{33}(\bQ,\omega)$, we obtain 
\begin{equation}
\chit_{0i}^{3a}(\bQ,\omega) \propto \omega^2
\end{equation}
for $a \in \{0,+,-\}$ and small $\omega$.
The product of imaginary and real parts of off-diagonal bare susceptibilities is thus of order $\omega^3$.
Combining all terms, we have thus shown that the out-of-plane damping term at $\bq=\bQ$ obeys
\begin{equation} \label{eq:damping3}
 {\rm Im} \frac{m^2}{\chit^{33}(\bQ,\omega)} \propto \omega^3
\end{equation}
at low frequencies.

For $\bq \neq \bQ$, the coherence factors remain finite at the hot spots (now determined by the equations $E_{\bk_H}^\ell = E_{\bk_H+\bq}^{\ell'} = 0$) so that
\begin{equation}
\chit_{0i}^{3a}(\bq,\omega) = - p^{3a}(\bq) \, \omega
\end{equation}
for small $\omega$ and $a \in \{0,+,-,3\}$.
However, the prefactor of this linear frequency dependence vanishes as $\bq$ approaches the ordering wave vector $\bQ$.
For the diagonal intraband coherence factor $A_{\ell\ell}^{33}(\bk,\bq)$, also the gradient with respect to $\bq$ vanishes at $\bk=\bk_H$ and $\bq=\bQ$.
Hence $p^{33}(\bq)$ is of order $(\bq-\bQ)^2$ for $\bq \to \bQ$.
For $a \neq 3$, the gradient $\nabla_\bq A_{\ell\ell}^{3a}(\bk,\bq)$ is finite at $\bk=\bk_H$ and $\bq=\bQ$, so that $p^{3a}(\bq)$ is of order $|\bq-\bQ|$ for $\bq \to \bQ$.
Eq.~\eqref{eq:chit33inv} can be generalized in the same form for $\bq \neq \bQ$.
For $\omega \to 0$ the contribution from $\chit_{0i}^{33}(\bq,\omega)$ is leading and yields
\begin{equation} \label{eq:damping3a}
 {\rm Im} \frac{m^2}{\chit^{33}(\bq,\omega)} = - \gamma(\bq) \, \omega + \cO(\omega^2) \, ,
\end{equation}
where $\gamma(\bq) \propto (\bq-\bQ)^2$ for $\bq \to \bQ$. The off-diagonal contributions to the damping term are of order $\omega^2$ for $\bq \neq \bQ$, with a prefactor that is linear in $|\bq-\bQ|$. Taking the limit $\omega \to 0$, $\bq \to \bQ$ at a fixed ratio $\hat\omega = \omega/|\bq-\bQ|$, diagonal and off-diagonal contributions are both of order $|\bq-\bQ|^3$.
The Landau damping of out-of-plane Goldstone modes thus scales to zero more rapidly than their excitation energy, so that these modes remain asymptotically stable quasi-particles.

The above results for the Landau damping hinge on the existence of hot spots. If Eq.~\eqref{eq:hotspots1} has no solution, the imaginary parts of the RPA susceptibilities are strictly zero below a certain threshold frequency. Higher order terms beyond RPA, such as fermionic self-energy contributions, will however yield a small low-frequency damping in any case.

Although electron and hole pockets coexist in the Brillouin zone for certain model parameters, we have not found any {\em interband}\/ hot spots in spiral states for the two-dimensional Hubbard model. If interband hot spots existed in a suitable system, an exceptionally large Landau damping would follow. Since the interband coherence factor $A_{\ell,-\ell}^{33}(\bk,\bQ)$ remains finite at the interband hot spots, the Landau damping term would be linear in $\omega$ even at $\bq=\bQ$, leading to a strong overdamping of the out-of-plane Goldstone mode.


\subsection{Special case: N\'eel state}
\label{sec:Neel}

The N\'eel state can be viewed as a special case of the spiral state where the ordering wave vector $\bQ$ assumes the special value $\bQ = (\pi,\pi)$ in two dimensions and $\bQ = (\pi,\pi,\pi)$ in three dimensions. In this section we analyze how the properties of the Goldstone modes derived above change in this case. In particular, we will see that the number of Goldstone modes is reduced to two, and their properties are equivalent.

The special properties of the N\'eel state are due to the fact that $\bQ$ and $-\bQ$ are identical wave vectors in the Brillouin zone if all components of $\bQ$ are equal to $\pi$. In other words $2\bQ$ is identical to $\bzero$.
As a first consequence, in the relation between the physical susceptibilities $\chi^{11}$ and $\chi^{22}$ and the susceptibilities $\chit^{ab}$ in the rotated spin basis, see Eq.~\eqref{eq:chi11phys}, terms which in the spiral state contribute only to off-diagonal (in momentum) susceptibilities $\chi^{aa}(\bq \pm 2\bQ,\bq,\omega)$ contribute to the momentum diagonal susceptibilities $\chi^{aa}(\bq,\bq,\omega)$ in the N\'eel state. Hence, instead of Eq.~\eqref{eq:chi11phys} one obtains
\begin{eqnarray}
 \chi^{11}(\bq,\omega) &=& \chit^{11}(\bq\pm\bQ,\omega) \, , \\
 \chi^{22}(\bq,\omega) &=& \chit^{22}(\bq\pm\bQ,\omega) \, .
\end{eqnarray}

In the spiral state we found three distinct Goldstone modes, an in-plane mode associated with a divergence of $\chit^{22}(\bq,\omega)$ for $\bq \to \bzero$ and $\omega \to 0$, and two out-of-plane modes leading to divergencies of $\chit^{33}(\bq,\omega)$ for $\bq \to \pm\bQ$ and $\omega \to 0$.
In the N\'eel state the two singularities of $\chit^{33}(\bq,\omega)$ collapse to one, since $\bQ$ and $-\bQ$ are now identical. Hence, only two Goldstone modes survive. This is in agreement with the fact that in the N\'eel state the continuous SU(2) spin rotation invariance is not completely broken: an U(1) symmetry associated with rotations around the spin orientation axis remains.
Moreover, in the N\'eel state the notion of ``in-plane'' and ``out-of-plane'' modes is meaningless since the N\'eel order singles out a particular axis, not a plane. The two Goldstone modes correspond to fluctuations of that axis in two orthogonal directions. By symmetry they must have the same stiffness, spectral weight and damping. We will now show that the properties of the in-plane and out-of-plane modes derived in the preceding section are indeed degenerate in the N\'eel limit.

In Appendix \ref{app:neel} we show that the off-diagonal bare susceptibilities $\chit_0^{02}$, $\chit_0^{03}$, $\chit_0^{12}$, and $\chit_0^{13}$ vanish identically in the N\'eel state. Hence, the sectors 0 and 1 are completely decoupled from the sectors 2 and 3 for all momenta and frequencies.
The expression \eqref{eq:J2RPA} for the in-plane stiffness thus simplifies to
\begin{equation}
 J_{\alpha\beta}^{(2)} = - 2 \Delta^2 \partial_{q_\alpha q_\beta}^2
 \chit_0^{22}(\bq,0) \big|_{\bq=\bzero} \, .
\end{equation}
Comparing with Eq.~\eqref{eq:J3RPA} for the out-of-plane stiffness, and using the relation $\chit_0^{22}(\bq,\omega) = \chit_0^{33}(\bq+\bQ)$ derived in Appendix~\ref{app:neel}, one finds $J_{\alpha\beta}^{(2)} = J_{\alpha\beta}^{(3)}$ as expected.

Due to the decoupling of the sectors 0 and 1 from the sectors 2 and 3, one can write $Z^{(3)}$ in a form analogous to the expression \eqref{eq:Z2RPA}, that is,
\begin{equation}
 Z^{(3)} = 2\Delta^2 \left[
 \partial_\omega^2 \chit_0^{33}(\bQ,\omega) \big|_{\omega=0} 
 + \frac{4U}{1 - 2U\chit_0^{22}(\bQ,0)}
 \left|\partial_\omega\chit_0^{23}(\bQ,\omega)\big|_{\omega=0}\right|^2
 \right] \, .
\end{equation}
Using once again $\chit_0^{22}(\bq,\omega) = \chit_0^{33}(\bq+\bQ)$, and $\chit_0^{23}(\bq,\omega) = \chit_0^{23}(\bq+\bQ)$ derived in Appendix~\ref{app:neel}, one obtains $Z^{(2)} = Z^{(3)}$.

We finally turn to the damping terms. In the N\'eel state, the intraband coherence factor $A_{\ell\ell}^{23}(\bk-\bq/2,\bq)$ is not only suppressed (of order $|\bq|^2$) for small $\bq$, but also for $\bq \to \bQ$, where it is of order $|\bq-\bQ|^2$. Combining this with the decoupling of the sectors 0 and 1 from the sectors 2 and 3, one obtains
\begin{equation}
 {\rm Im} \, \frac{1}{\chit^{22}(\bq,\omega)} = - 4U^2
 \chit_{0i}^{22}(\bq,\omega) + {\cal O}(|\bq|^3)
\end{equation}
for small $\bq$, and
\begin{equation}
 {\rm Im} \, \frac{1}{\chit^{33}(\bq,\omega)} = - 4U^2
 \chit_{0i}^{33}(\bq,\omega) + {\cal O}(|\bq-\bQ|^3)
\end{equation}
for small $\bq-\bQ$. The relation $\chit_0^{22}(\bq,\omega) = \chit_0^{33}(\bq+\bQ)$ then implies that the damping of the 2-mode and the 3-mode is identical.
Returning to the susceptibilities in the physical (unrotated) spin basis one obtains
\begin{equation}
 {\rm Im} \, \frac{m^2}{\chi^{22}(\bq,\omega)} =
 {\rm Im} \, \frac{m^2}{\chi^{33}(\bq,\omega)} =
 - |\bq'|^2 \gamma(\hat\bq',\hat\omega) + \cO(|\bq'|^3)
\end{equation}
for small $\bq' = \bq - \bQ$ and fixed $\hat\omega = \omega/|\bq'|$.
This form of the Landau damping in a N\'eel state has already been derived by Sachdev et al.~\cite{sachdev95}.


\subsection{Numerical results in two dimensions}

To complement our general results, and to get an idea about the typical size of the spin stiffnesses and the damping terms, we now present some numerical results as obtained by evaluating the analytic expressions derived above for a specific model in two dimensions: the repulsive Hubbard model on the square lattice with nearest and next-to-nearest neighbor hopping amplitudes ($t$ and $t'$, respectively). We choose $t$ as our unit of energy, that is, all results with an energy dimension are presented for $t=1$.

\begin{figure}
\centering
\includegraphics[width=9cm]{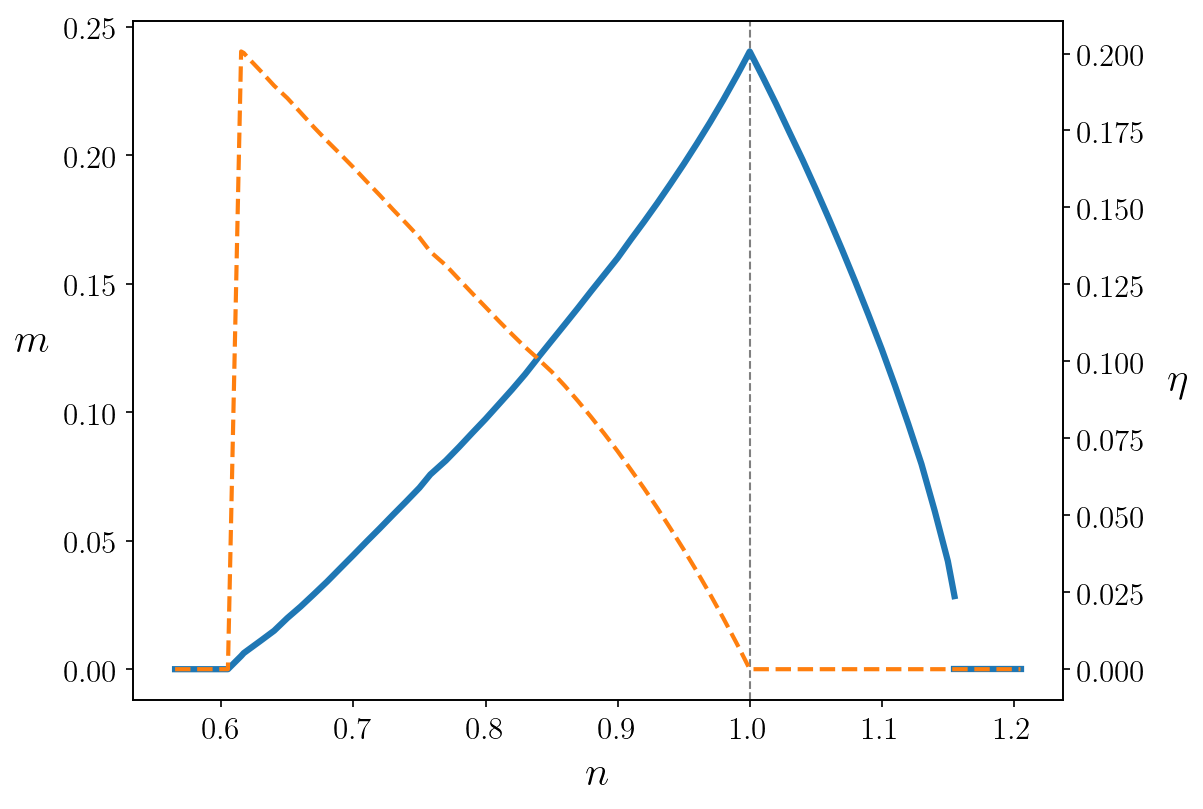}
\caption{Magnetization $m$ (left axis, solid line) and incommensurability $\eta$ (right axis, dashed line) as a function of the electron density $n$ in the mean-field ground state of the two-dimensional Hubbard model with parameters $t'/t = -0.16$ and $U/t = 2.5$.}
\label{fig:meta}
\end{figure}
We compute only ground state properties. We choose $t' = -0.16t$ and a relatively weak Hubbard interaction $U = 2.5t$. For this choice of parameters mean-field theory yields a homogeneous spiral magnetic state over an extended density range between $n \approx 0.61$ and $n = 1$ (half-filling). At half-filling and for electron doping up to $n \approx 1.15$ the simple N\'eel state minimizes the mean-field energy. In the spiral state for $n < 1$ the ordering wave vector has the form $\bQ = (\pi - 2\pi\eta,\pi)$. The incommensurability $\eta$ increases monotonically upon reducing the density, and vanishes continuously for $n \to 1$. The onset of the spiral order at $n \approx 0.61$ is continuous, while the transition between the N\'eel state and the paramagnetic state at $n \approx 1.15$ is of first order, albeit with a relatively small jump of the order parameter. The magnetization $m$ and the incommensurability $\eta$ are plotted as functions of the electron density $n$ in Fig.~\ref{fig:meta}.

\begin{figure}
\centering
\includegraphics[width=9cm]{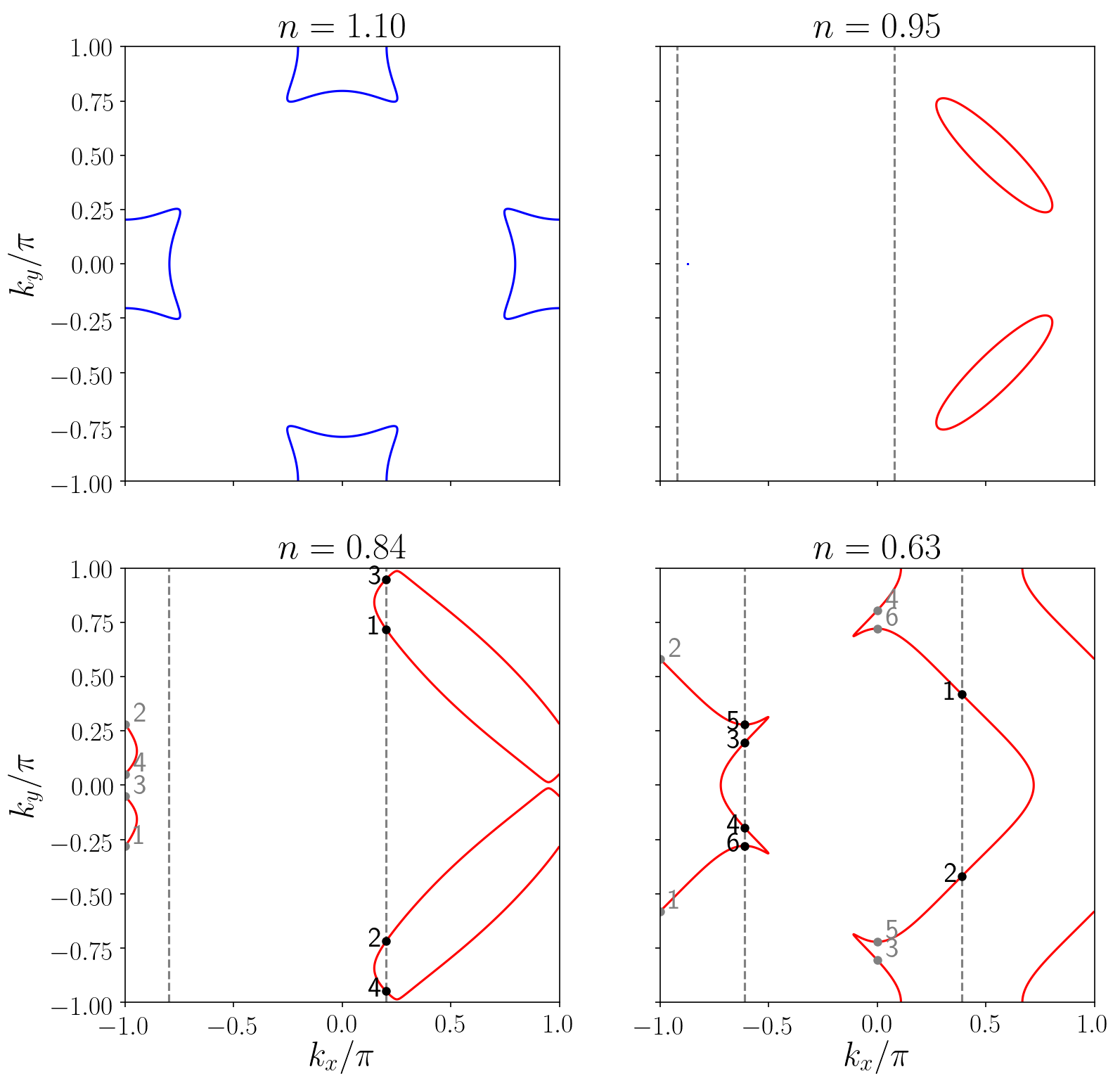}
\caption{Quasiparticle Fermi surfaces in the magnetic ground state at various electron densities. Blue lines correspond to momenta satisfying $E_{\bk}^+ = 0$, red lines to momenta satisfying $E_{\bk}^- = 0$. The dashed vertical lines at $k_x = 2\pi\eta$ and $k_x = 2\pi\eta - \pi$ are solutions of the equation $\xi_{\bk+2\bQ} = \xi_\bk$. For $n=0.84$ and $n=0.63$ there are hot spots on the Fermi surfaces (black dots) which are connected to other points on the Fermi surfaces (grey dots) by a momentum shift $\bQ$. The numbers indicate the pairwise connection. In the N\'eel state at $n=1.1$ all points on the Fermi surface are connected to each other by $\bQ = (\pi,\pi)$.}
\label{fig:FS}
\end{figure}
In Fig.~\ref{fig:FS} we show the quasiparticle Fermi surfaces in the magnetic ground state at various electron densities from $n=0.63$ to $n=1.1$. For $n<1$ these are given by momenta satisfying the equation $E_{\bk}^- = 0$, for $n>1$ by solutions of $E_{\bk}^+ = 0$. In the spiral state for $n<1$ hot spots corresponding to solutions of Eqs.~\eqref{eq:hotspots1} or \eqref{eq:hotspots2} exist only for sufficiently large hole doping at $n=0.84$ and $n=0.63$. Hence, for low hole doping, such as $n=0.95$, there is no Landau damping of the out-of-plane magnons. 

\begin{figure}
\centering
\includegraphics[width=9cm]{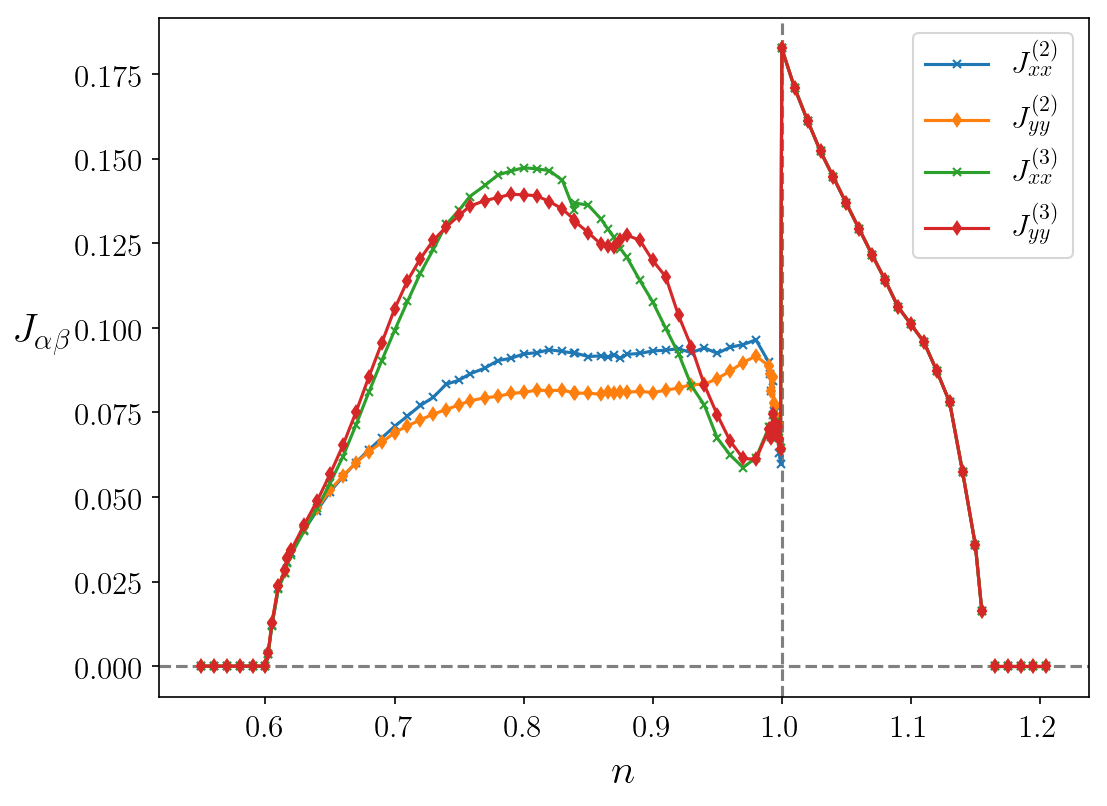}
\caption{In-plane and out-of-plane spin stiffnesses as a function of the electron density. In the N\'eel state for $n \geq 1$ all stiffnesses assume the same value.}
\label{fig:Jab}
\end{figure}
In Fig.~\ref{fig:Jab} we show the in-plane and out-of-plane spin stiffnesses $J_{\alpha\beta}^{(a)}$ as a function of the electron density. Both in the spiral state for $n < 1$ and in the N\'eel state for $n \geq 1$ only diagonal components $J_{\alpha\alpha}^{(a)}$ with $\alpha = x,y$ are non-zero. In the N\'eel state the stiffnesses are isotropic (independent of $\alpha$) and degenerate ($J_{\alpha\alpha}^{(2)} = J_{\alpha\alpha}^{(3)}$), as dictated by symmetry.
The spin stiffnesses are positive for all densities where a magnetic solution exists, showing that the spiral state for $n<1$ and the N\'eel state for $n \geq 1$ are at least meta stable.
In the spiral state the in-plane and out-of-plane stiffnesses differ significantly among each other, except for the lowest densities (where $m \to 0$) and near half-filling. Both exhibit a slight nematicity (dependence on $\alpha$) which comes from the difference between $Q_x$ and $Q_y$. All spin stiffnesses $J_{\alpha\alpha}^{(a)}$ exhibits a pronounced jump at half-filling. More precisely, upon approaching half-filling from below ($n<1$) the stiffnesses converge to a value that differs from $J_{\alpha\alpha}^{(a)}$ at half-filling. This discontinuity is caused by the sudden appearance of hole-pockets upon hole-doping, which allow for intraband processes with small excitation energies. A discontinuity due to electron pockets upon approaching half-filling from above ($n>1$) is prevented by vanishing prefactors at the momenta $(\pi,0)$ and $(0,\pi)$ where the electron pockets pop up.

\begin{figure}
\centering
\includegraphics[width=9cm]{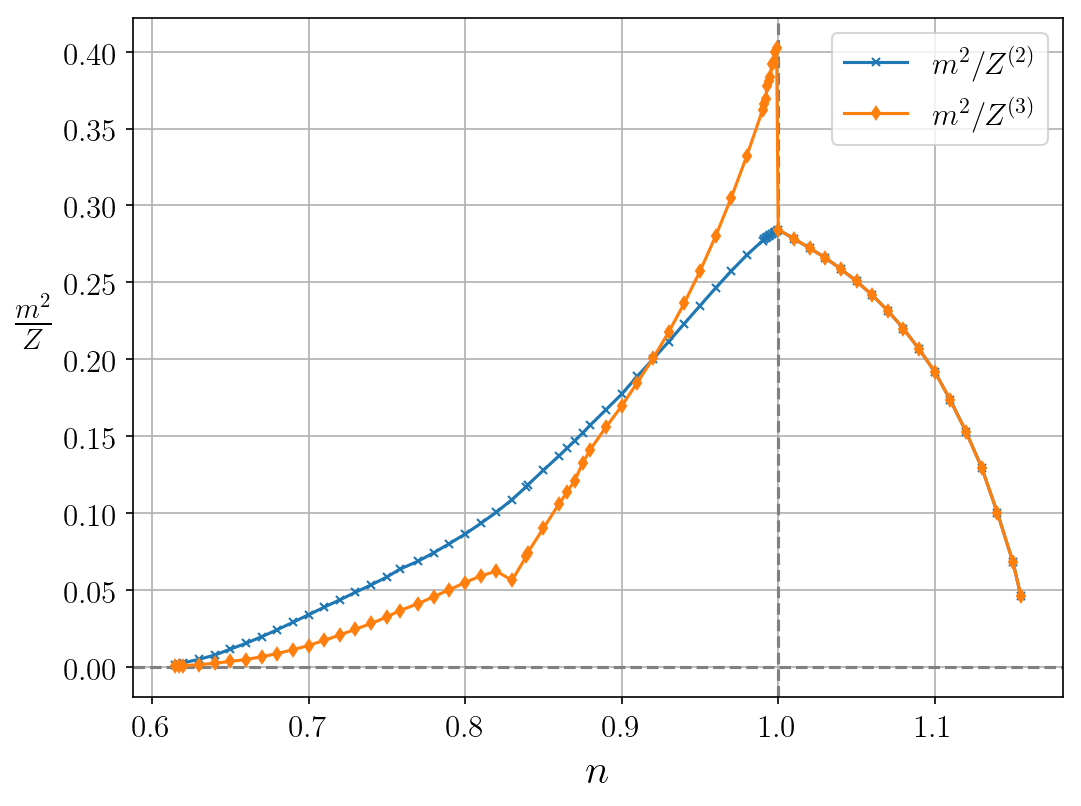}
\caption{In-plane and out-of-plane spectral weights as a function of the electron density. In the N\'eel state for $n \geq 1$ both weights assume the same value.}
\label{fig:specweight}
\end{figure}
The density dependence of the spectral weights of the magnon modes $m^2/Z^{(a)}$ is shown in Fig.~\ref{fig:specweight}. In the spiral state for $n < 1$ there is a pronounced difference between the in-plane and the out-of-plane modes. The discontinuity of $m^2/Z^{(3)}$ at half-filling is again due to intraband contributions within the hole pockets emerging for $n < 1$. By contrast, $m^2/Z^{(2)}$ is continuous, since only interband terms contribute to the in-plane $Z$-factor. Both spectral weights are positive in the entire ordered phase. The spectral weights decrease near the edges of the magnetic regime, since $m^2$ vanishes more rapidly than $Z^{(a)}$ upon approaching the edges. The weight of the out-of-plane mode $m^2/Z^{(3)}$ exhibits a dip at the density $n \approx 0.84$ where two hole pockets merge. 

\begin{figure}
\centering
\includegraphics[width=9cm]{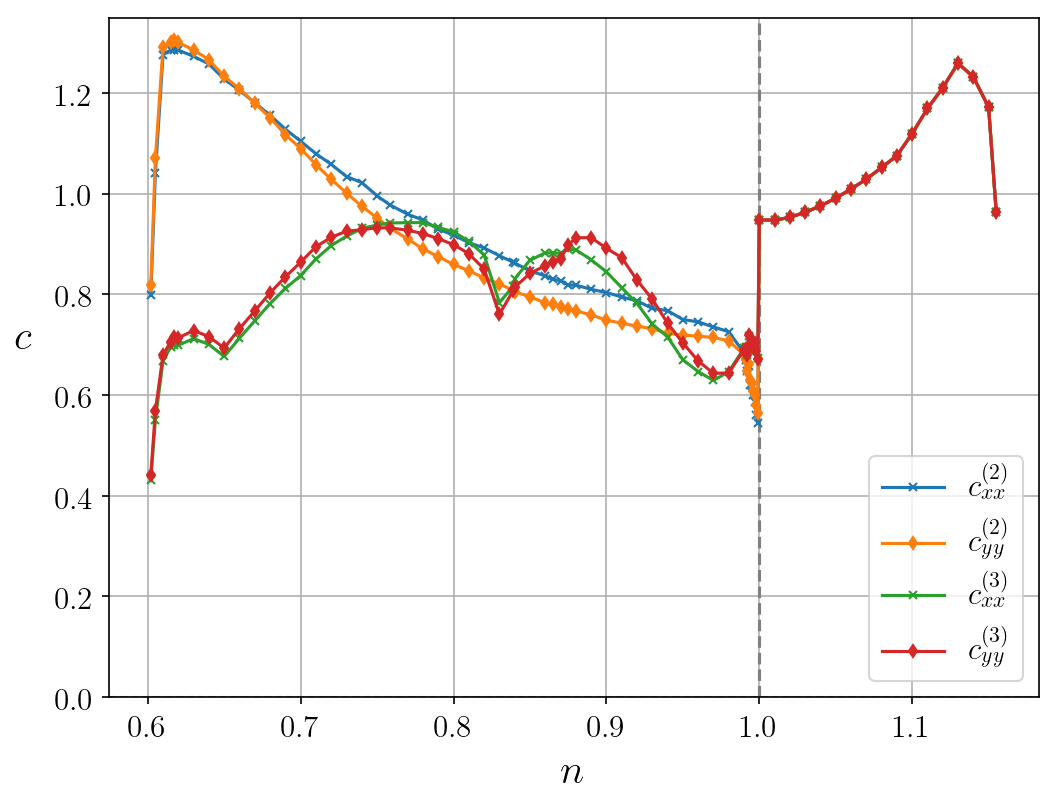}
\caption{In-plane and out-of-plane magnon velocities $c_{\alpha\alpha}^{(a)} = \big[J_{\alpha\alpha}^{(a)}/Z^{(a)}\big]^{1/2}$ as a function of the electron density.}
\label{fig:velocities}
\end{figure}
In Fig.~\ref{fig:velocities} we show the magnon velocities $c_{\alpha\alpha}^{(a)} = \big[J_{\alpha\alpha}^{(a)}/Z^{(a)}\big]^{1/2}$. The velocities exhibit only a moderate density dependence. Their size is over order one (in units of $t$) in the entire magnetic regime.

\begin{figure}
\centering
\includegraphics[width=9cm]{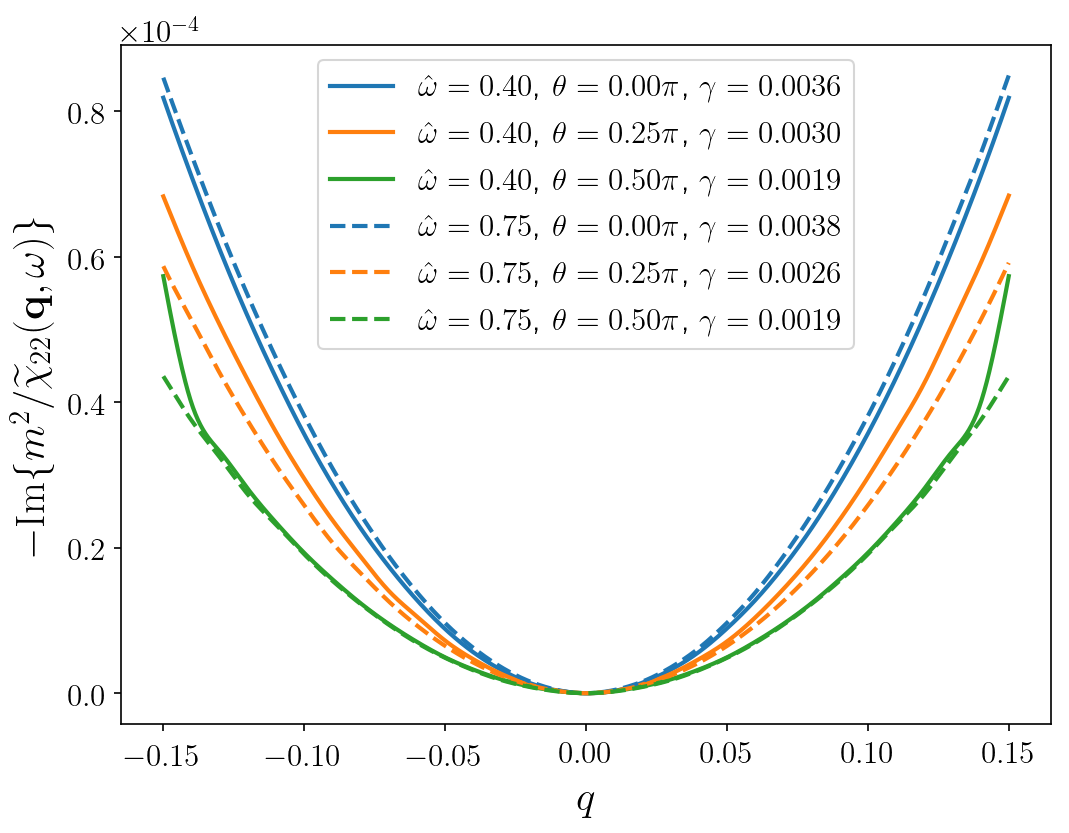}
\caption{Damping term of the in-plane Goldstone mode as a function of $|\bq|$ for two fixed values of $\hat\omega$ and a density $n=0.84$. Various directions of $\bq$ are parametrized by the angle $\theta$ between $\bq$ and the $q_x$-axis. The prefactor $\gamma$ of the leading quadratic dependence on $|\bq|$ is shown in the inset.}
\label{fig:damp2}
\end{figure}
In Fig.~\ref{fig:damp2} we plot the in-plane damping term ${\rm Im} [m^2/\chit^{22}(\bq,\omega)]$ as a function of $|\bq|$ at two fixed values of $\hat\omega = \omega/|\bq|$ and three fixed directions $\hat\bq = \bq/|\bq|$. The density is fixed at $n=0.84$. One can see the quadratic dependence on $|\bq|$ in agreement with Eq.~\eqref{eq:damping2}. The prefactors $\gamma(\hat\omega,\hat\bq)$ are shown in the inset.

\begin{figure}
\centering
\includegraphics[width=9cm]{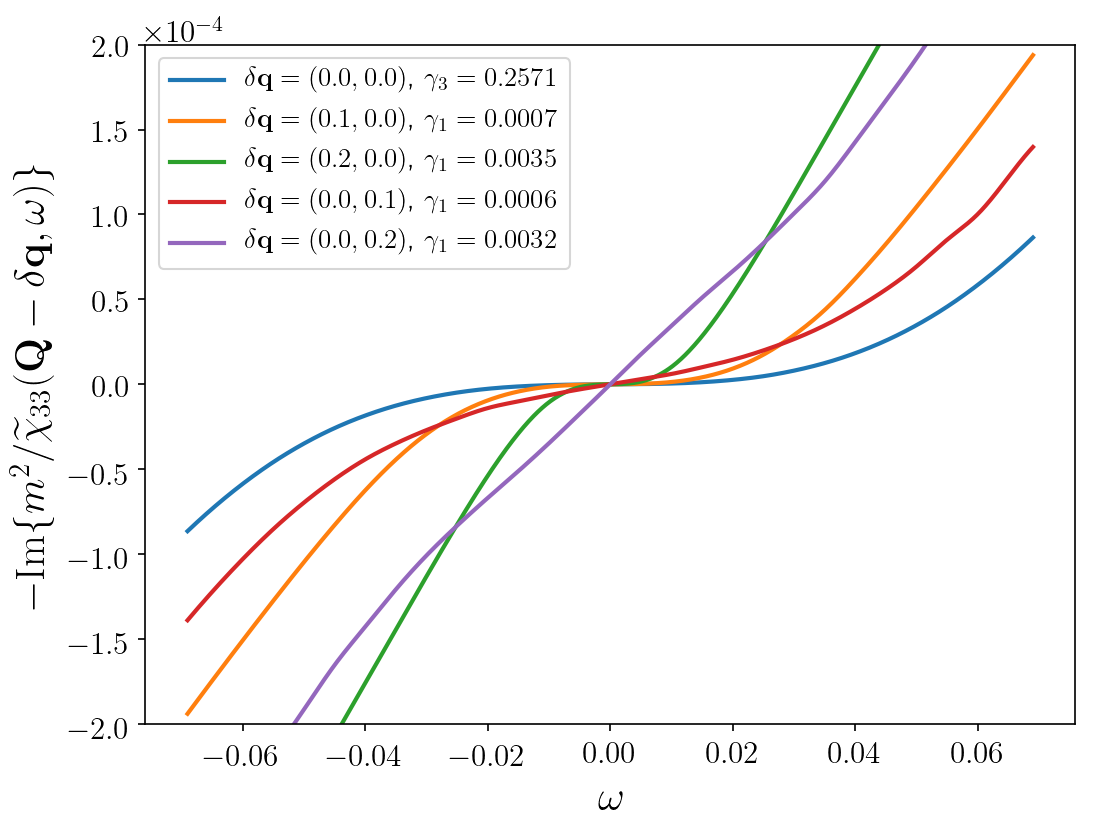}
\caption{Damping term of the out-of-plane Goldstone mode as a function of $\omega$ for various fixed wave vectors $\bq$ near $\bQ = (0.82\pi,\pi)$ and fixed density $n=0.84$. The prefactors $\gamma_1$ of the linear frequency dependence for $\bq \neq \bQ$ and the prefactor $\gamma_3$ of the cubic frequency dependence for $\bq = \bQ$ are shown in the inset.}
\label{fig:damp3}
\end{figure}
The frequency dependence of the out-of-plane damping ${\rm Im} [m^2/\chit^{33}(\bq,\omega)]$ is shown in Fig.~\ref{fig:damp3} for various fixed momenta $\bq$ at and near $\bQ$. For $\bq = \bQ$ the damping is proportional to $\omega^3$ for low frequencies, in agreement with Eq.~\eqref{eq:damping3}. For $\bq \neq \bQ$ one can see the linear frequency dependence in agreement with Eq.~\eqref{eq:damping3a}. The prefactors of the leading cubic and linear terms are listed in the inset.


\section{Conclusion}

In summary, we have investigated the properties of the Goldstone modes (that is, magnons) in metallic electron systems with spiral magnetic order. Our analysis is based on the RPA susceptibilities of tight binding electrons with an arbitrary dispersion and a local Hubbard interaction. In agreement with  general arguments and previous studies \cite{rastelli85, chandra90, shraiman92, kampf96} we have identified three Goldstone poles in the susceptibilities, one associated with in-plane, and two associated with out-of-plane fluctuations of the order parameter. The energy-momentum relations of all the modes are linear.

We have derived expressions for the spin stiffnesses and the spectral weights of the magnons, from which the magnon velocities can be obtained, too. The expressions for the spin stiffnesses are also useful for checking the stability of the spiral state, for example, against an out-of-plane canting of the spins. Moreover, we have determined the size of the decay rates of the magnons due to Landau damping. The Landau damping of the in-plane mode has the same form as for the Goldstone modes in a N\'eel antiferromagnet \cite{sachdev95} and is of the same order as the energy $\omega$ of the mode. By contrast, the Landau damping of the out-of-plane modes is smaller, of the order $\omega^{3/2}$. Hence, the out-of-plane modes are asymptotically stable excitations in the low energy limit.

We have complemented our general analysis with a numerical evaluation of the spin stiffnesses, spectral weights, and decay rates for a specific two-dimensional model system. Some of the quantities exhibit peaks and discontinuities as a function of the electron density which are related to changes of the Fermi surface topology and special contributions in the N\'eel state.

Magnons and their decay rates can in principle be detected by inelastic neutron scattering. Our analysis indicates that out-of-plane magnon branches in a metallic spiral magnet should be sharper than the in-plane branch at low excitation energies.


\begin{acknowledgments}
We are grateful to A.~Chubukov, L.~Classen, L.~Debbeler, B.~Keimer, E.~K\"onig, J.~Mitscherling, O.~Sushkov, J.~Sykora, and D.~Vilardi for valuable discussions.
\end{acknowledgments}

\vskip 5mm


\begin{appendix}


\section{Coherence factors}
\label{app:cohfac}
The coherence factors entering the bare susceptibilities $\chit_0^{ab}$ in Eq.~\eqref{eq:chit0} are defined as
\begin{equation}
 A^{ab}_{\ell\ell'}(\bk,\bq) =
 \frac{1}{2} \tr\left[ \sigma^a\,u_{\bk}^\ell\,\sigma^b\,u_{\bk+\bq}^{\ell'}
 \right] \, ,
\end{equation}
where $\ell,\ell'$ are the quasi-particle band indices, $a \in \{ 0,1,2,3 \}$ labels the charge and spin components, and the functions $u_\bk^\ell$ are the linear combinations of Pauli matrices defined in Eq.~\eqref{eq:ukell}.
Performing the trace we obtain explicit expressions.
For the charge-charge coherence factor we get
\begin{equation} \label{eq:A00}
 A^{00}_{\ell\ell'}(\bk,\bq) =
 1 + \ell\ell'\,\frac{h_\bk h_{\bk+\bq} + \Delta^2}{e_\bk e_{\bk+\bq}} \, ,
\end{equation}
while for the charge-spin ones we find
\begin{eqnarray}
\label{eq:A01}
 A^{01}_{\ell\ell'}(\bk,\bq) &=&
 \ell \, \frac{\Delta}{e_\bk} + \ell' \, \frac{\Delta}{e_{\bk+\bq}} \, , \\
\label{eq:A02}
 A^{02}_{\ell\ell'}(\bk,\bq) &=&
 i \ell\ell' \, \Delta \frac{h_\bk - h_{\bk+\bq}}{e_\bk e_{\bk+\bq}} \, , \\
\label{eq:A03}
 A^{03}_{\ell\ell'}(\bk,\bq) &=&
 \ell \, \frac{h_\bk}{e_\bk} + \ell' \, \frac{h_{\bk+\bq}}{e_{\bk+\bq}} \, .
\end{eqnarray}
The diagonal coherence factors in the spin subsector are given by
\begin{eqnarray}
\label{eq:A11}
 A^{11}_{\ell\ell'}(\bk,\bq) &=&
 1 - \ell\ell'\,\frac{h_\bk h_{\bk+\bq} - \Delta^2}{e_\bk e_{\bk+\bq}} \, , \\
\label{eq:A22}
 A^{22}_{\ell\ell'}(\bk,\bq) &=&
 1 - \ell\ell'\,\frac{h_\bk h_{\bk+\bq} + \Delta^2}{e_\bk e_{\bk+\bq}} \, , \\
\label{eq:A33}
 A^{33}_{\ell\ell'}(\bk,\bq) &=&
 1 + \ell\ell'\,\frac{h_\bk h_{\bk+\bq} - \Delta^2}{e_\bk e_{\bk+\bq}} \, ,
\end{eqnarray}
and the off-diagonal ones by
\begin{eqnarray}
\label{eq:A12}
 A^{12}_{\ell\ell'}(\bk,\bq) &=&
 i\ell \, \frac{h_\bk}{e_\bk} - i\ell' \, \frac{h_{\bk+\bq}}{e_{\bk+\bq}} \, , \\
\label{eq:A13}
 A^{13}_{\ell\ell'}(\bk,\bq) &=&
 \ell\ell' \, \Delta \frac{h_\bk + h_{\bk+\bq}}{e_\bk e_{\bk+\bq}} \, , \\
\label{eq:A23}
 A^{23}_{\ell\ell'}(\bk,\bq) &=&
 i \ell \, \frac{\Delta}{e_\bk} - i\ell' \, \frac{\Delta}{e_{\bk+\bq}} \, .
\end{eqnarray}
The coherence factors for $a>b$ are obtained from the general relation
$A^{ab}_{\ell\ell'}(\bk,\bq) = \big[ A^{ba}_{\ell\ell'}(\bk,\bq) \big]^*$.
The coherence factors are purely imaginary if (and only if) exactly one of the indices $a,b$ is equal to two, and they are real otherwise.
Hence, the exchange of the indices $a$ and $b$ yields
\begin{equation} \label{eq:Aabexchange}
 A_{\ell\ell'}^{ba}(\bk,\bq) = p^a p^b A_{\ell\ell'}^{ab}(\bk,\bq),
\end{equation}
where $p^a=+1$ for $a=0,1,3$, and $p^a=-1$ for $a=2$.

From $\xi_\bk=\xi_{\mbk}$ one obtains the relations
$h_{-\bk-\bQ} = -h_\bk$,
$g_{-\bk-\bQ} = g_\bk$,
$e_{-\bk-\bQ} = e_\bk$, and
$u^\ell_{-\bk-\bQ} = \sigma^1\,u_\bk^\ell\,\sigma^1$.
From Eq.~\eqref{eq:Acalc} we then see that
\begin{equation}
 A^{ab}_{\ell'\ell}(\mbk-\bQ-\bq,\bq) =
 \frac{1}{2} \tr\left[\tilde{\sigma}^b\, u^{\ell}_{\bk+\bq} \,
 \tilde{\sigma}^a\,u^{\ell'}_{\bk}\right] \, ,
\end{equation}
with $\tilde{\sigma}^a=\sigma^1\sigma^a\sigma^1=s^a\sigma^a$, where $s^a=+1$ for $a=0,1$, and $s^a=-1$ for $a=2,3$. Using Eq.~\eqref{eq:Aabexchange}, we then obtain
\begin{equation} \label{eq:Asymmetry}
 A^{ab}_{\ell'\ell}(\mbk-\bQ-\bq,\bq) = 
 s^a s^b A^{ba}_{\ell\ell'}(\bk,\bq) = 
 s^{ab} \, A^{ab}_{\ell\ell'}(\bk,\bq) \, ,
\end{equation}
where
\begin{equation} \label{eq:sab}
 s^{ab} = s^a s^b p^a p^b = (1-2\delta_{a3})(1-2\delta_{b3}) \, .
\end{equation}
The relation \eqref{eq:Asymmetry} will be useful in the following section.


\section{Symmetries of the bare susceptibilities}
\label{app:chi0symm}

In this appendix we derive the behavior of the bare susceptibilities under sign changes of the frequency and the momentum arguments.


\subsection{Parity under frequency sign change}

We decompose the expression (\ref{eq:chit0}) for the susceptibility components in intraband and interband contributions 
\begin{equation}
 \chit^{ab}_0(\bq,\omega) =
 -\frac{1}{8}\sum_\ell\int_\bk A^{ab}_{\ell\ell}(\bk,\bq)\frac{f(E^\ell_\bk)-f(E^\ell_\bkq)}{E^\ell_\bk-E^\ell_\bkq+z}
 -\frac{1}{8}\sum_\ell\int_\bk A^{ab}_{\ell,\ellnot}(\bk,\bq)\frac{f(E^\ell_\bk)-f(E^\ellnot_\bkq)}{E^\ell_\bk-E^\ellnot_\bkq+z} \, ,
\end{equation}
where $z=\omega+i0^+$.
Substituting $\bk \to \mbk-\bQ-\bq$, the intraband term can be rewritten as
\begin{eqnarray}
 [\chit^{ab}_0(\bq,\omega)]_\mathrm{intra} &=&
 -\frac{1}{8} \sum_\ell\int_\bk A^{ab}_{\ell\ell}(\bk,\bq)
 \frac{f(E^\ell_\bk)}{E^\ell_\bk-E^\ell_\bkq+z} \nonumber \\
 && -\frac{1}{8} \sum_\ell\int_\bk A^{ab}_{\ell\ell}(\mbk-\bQ-\bq,\bq)
 \frac{-f(E^\ell_{\mbk-\bQ})}{-(E^\ell_{\mbk-\bQ}-E^\ell_{\mbk-\bQ-\bq}-z)} \, .
\end{eqnarray}
Using Eq.~\eqref{eq:Asymmetry} and $E^\ell_{\mbk-\bQ} = E^\ell_{\bk}$, we obtain
\begin{equation}
 [\chit^{ab}_0(\bq,\omega)]_\mathrm{intra} =
 -\frac{1}{8} \sum_\ell\int_\bk A^{ab}_{\ell\ell}(\bk,\bq) f(E^\ell_\bk)
 \left( \frac{1}{E^\ell_\bk-E^\ell_\bkq+z} +
 \frac{s^{ab}}{E^\ell_\bk-E^\ell_\bkq-z} \right) \, .
\end{equation}
Similarly, the interband term can be rewritten as
\begin{eqnarray}
 [\chit^{ab}_0(\bq,\omega)]_\mathrm{inter} &=&
 -\frac{1}{8}\sum_\ell\int_\bk A^{ab}_{\ell,\ellnot}(\bk,\bq)
  \frac{f(E^\ell_\bk)}{E^\ell_\bk-E^\ellnot_\bkq+z} \nonumber \\
 && -\frac{1}{8}\sum_\ell\int_\bk A^{ab}_{\ellnot,\ell}(\mbk- \bq-\bQ,\bq)
 \frac{-f(E^\ell_{\mbk-\bQ})}{-(E^\ell_{\mbk-\bQ}-E^\ellnot_{\mbk-\bQ-\bq}-z)} \, .
\end{eqnarray}
In the second term we have also made the substitution $\ell \to -\ell$.
Using Eq.~(\ref{eq:Asymmetry}) for $\ell'=-\ell$, we get
\begin{equation}
 [\chit^{ab}_0(\bq,\omega)]_\mathrm{inter} =
 -\frac{1}{8}\sum_\ell\int_\bk A^{ab}_{\ell,\ellnot}(\bk,\bq) f(E^\ell_\bk)
 \left( \frac{1}{E^\ell_\bk-E^\ellnot_\bkq+z} + 
 \frac{s^{ab}}{E^\ell_\bk-E^\ellnot_\bkq-z} \right) \, ,
\end{equation}
with $s^{ab}$ as defined in Eq.~\eqref{eq:sab}.
Summing the intraband and the interband terms we obtain
\begin{eqnarray}
\label{eq:minusomega1}
 \chit_{0r}^{ab}(\bq,-\omega) &=& {\phantom -} s^{ab} \chit_{0r}^{ab}(\bq,\omega) \\
\label{eq:minusomega2}
 \chit_{0i}^{ab}(\bq,-\omega) &=& - s^{ab} \chit_{0i}^{ab}(\bq,\omega) \, .
\end{eqnarray}
%


\subsection{Parity under momentum sign change}

Substituting $\bk \to \bk - \bq/2$, we rewrite the bare susceptibility as
\begin{equation}
 \chit_0^{ab}(\bq,\omega)= -\frac{1}{8}\sum_{\ell\ell'} \int_\bk
 A^{ab}_{\ell\ell'}\left(\bk-\frac{\bq}{2},\bq\right)
 \frac{f(E^{\ell}_{\bk-\frac{\bq}{2}})-f(E^{\ell'}_{\bk+\frac{\bq}{2}})}
 {E^{\ell}_{\bk-\frac{\bq}{2}}-E^{\ell'}_{\bk+\frac{\bq}{2}}+\omega+i0^+} \, .
\end{equation}
Using
\begin{equation}
 A^{ab}_{\ell'\ell}\left(\bk+\frac{\bq}{2},-\bq\right) =
 A^{ba}_{\ell\ell'}\left(\bk-\frac{\bq}{2},\bq\right) = p^{a}p^{b}\,A^{ab}_{\ell\ell'}\left(\bk-\frac{\bq}{2},\bq\right) \, ,
\end{equation}
with $p^{a}$ as defined in Appendix~\ref{app:cohfac}, we immediately see that 
\begin{equation} \label{eq:Aq->-q}
 \chit_0^{ab}(-\bq,-\omega) = p^{a}p^{b} \, \chit_0^{ab}(\bq,\omega) \, .
\end{equation}
Combining this with Eqs.~\eqref{eq:minusomega1} and \eqref{eq:minusomega2}, we obtain
\begin{eqnarray}
\label{eq:minusq1}
 \chit_{0r}^{ab}(-\bq,\omega) &=& {\phantom -} p^{ab} \chit_{0r}^{ab}(\bq,\omega) \\
\label{eq:minusq2}
 \chit_{0i}^{ab}(-\bq,\omega) &=& - p^{ab} \chit_{0i}^{ab}(\bq,\omega) \, ,
\end{eqnarray}
\begin{equation}
 p^{ab} = p^{a} p^{b} s^{ab} = s^a s^b =
 (1-2\delta_{a2})(1-2\delta_{b2})(1-2\delta_{a3})(1-2\delta_{b3}) \, .
\end{equation}
%


\section{Calculation of $\chit_0^{33}(\pm\bQ,0)$}
\label{app:chi33}
In this Appendix we prove the relation \eqref{eq:chi33_0(q,0)} for $\chit_0^{33}(-\bQ,0)$. The corresponding relation for $\chit_0^{33}(\bQ,0)$ follows from the parity of $\chit_0^{33}(\bq,\omega)$ under $\bq\to-\bq$. Using the general expression~(\ref{eq:chit0}) for the bare susceptibility, and Eq.~\eqref{eq:A33} for the coherence factor $A^{33}_{\ell\ell'}(\bk,\bq)$, one obtains
\begin{eqnarray}
 \chit^{33}_0(-\bQ,0) &=&
 - \frac{1}{8} \int_\bk \left[ 1 + \frac{h_\bk h_{\bk-\bQ}-\Delta^2}{e_\bk e_{\bk-\bQ}}\right]
 \left(\frac{f(E^+_\bk)-f(E^+_{\bk-\bQ})}{E^+_\bk-E^+_{\bk-\bQ}} +
 \frac{f(E^-_\bk)-f(E^-_{\bk-\bQ})}{E^-_\bk-E^-_{\bk-\bQ}}\right) \nonumber \\
 && - \frac{1}{8} \int_\bk \left[1-\frac{h_\bk h_{\bk-\bQ}-\Delta^2}{e_\bk e_{\bk-\bQ}}\right]
 \left(\frac{f(E^+_\bk)-f(E^-_{\bk-\bQ})}{E^+_\bk-E^-_{\bk-\bQ}} +
 \frac{f(E^-_\bk)-f(E^+_{\bk-\bQ})}{E^-_\bk-E^+_{\bk-\bQ}}\right) \nonumber \\[2mm]
 &=& - \frac{1}{4} \sum_{\ell=\pm}\int_\bk 
 \left\{\left[1-\frac{h_\bk h_\mbk+\Delta^2}{e_\bk e_\mbk}\right]
 \frac{f(E^\ell_\bk)}{E^\ell_\bk-E^\ell_\mbk} +
 \left[1+\frac{h_\bk h_\mbk+\Delta^2}{e_\bk e_\mbk}\right]
 \frac{f(E^\ell_\bk)}{E^\ell_\bk-E^\ellnot_\mbk}\right\} \nonumber \\[2mm]
 &=& \sum_{\ell=\pm} \int_\bk \frac{(-\ell) f(E^\ell_\bk)}{4e_\bk}
 \left\{\frac{2\ell e_\bk (g_\bk-g_\mbk)+2 h_\bk(h_\bk -h_\mbk)}
 {(E^\ell_\bk-E^\ellnot_\mbk)(E^\ell_\bk-E^\ell_\mbk)}\right\} \, .
\end{eqnarray}
In the second equation we have used $h_{\bk-\bQ}=-h_{\mbk}$, $e_{\bk-\bQ}=e_{\mbk}$, and $E^\pm_{\bk-\bQ}=E^\pm_{\mbk}$. It is easy to see that the linear combinations
$\gmk = g_\bk - g_\mbk$, $h^\pm_\bk = h_\bk \pm h_\mbk$, and $e^\pm_\bk = e_\bk \pm e_\mbk$
obey the relations $\hmk\hpk = h_\bk^2-h_\mbk^2 = e_\bk^2-e_\mbk^2 = \emk\epk$, and
$\hmk = -\gmk$. Using these relations, we finally get
\begin{eqnarray}
 \chit_0^{33}(-\bQ,0) &=&
  \sum_{\ell=\pm}\int_\bk \frac{(-\ell) f(E^\ell_\bk)}{4e_\bk}
  \left\{\frac{2\ell e_\bk \gmk + 2h_\bk\hmk }{(\gmk + \ell \epk)(\gmk + \ell \emk)}
  \right\} \nonumber \\
  &=& \sum_{\ell=\pm}\int_\bk \frac{(-\ell) f(E^\ell_\bk)}{4e_\bk}
  \left\{\frac{2\ell e_\bk \gmk + \emk\epk + (\gmk)^2 }{(\gmk + \ell \epk)(\gmk + \ell \emk)}
  \right\} \nonumber \\
  &=& \sum_{\ell=\pm}\int_\bk \frac{(-\ell) f(E^\ell_\bk)}{4e_\bk} =
  \int_\bk \frac{f(E^-_\bk) - f(E^+_\bk)}{e_\bk} \, .
\end{eqnarray}
%


\section{Expansion of $1/\chit^{22}(\bq,0)$ for small $\bq$}
\label{app:chi22}

Here we derive the expansion of $1/\chit^{22}(\bq,0)$ for small $\bq$ to quadratic order by using a block form of the susceptibility matrix and Schur's complement. The expansion of $1/\chit^{33}(\bQ,\omega)$ for small $\omega$ proceeds in close analogy.

Since $\chit_0^{a3}(\bq,0) = \chit_0^{3a}(\bq,0) = 0$ for $a \neq 3$, the 3-component is decoupled from all the other components for $\omega = 0$, so that we need to consider only matrix elements with indices $0,1,2$. Hence, in this appendix, $\chit$, $\chit_0$ and $\Gamma_0$ denote $3 \times 3$ matrices formed only by these matrix elements.
We write $\chit_0$ and $\Gamma_0$ in block form
\begin{equation}
 \chit_0 = \left( \begin{array}{cc} 
 \bar\chit_0 & v \\ v^\dagger & \chit_0^{22} \end{array} \right) \, , \quad
 \Gamma_0 = \left( \begin{array}{cc} 
 \bar\Gamma_0 & 0 \\ 0 & 2U 
 \end{array} \right) \, ,
\end{equation}
where
\begin{equation}
 \bar\chit_0 = \left( \begin{array}{cc} 
 \chit_0^{00} & \chit_0^{01} \\ \chit_0^{10} & \chit_0^{11} 
 \end{array} \right) \, , \quad
 \bar\Gamma_0 = \left( \begin{array}{cc} 
 -2U & 0 \\ 0 & 2U 
 \end{array} \right) \, ,
\end{equation}
and
\begin{equation}
 v = \left( \begin{array}{c} \chit_0^{02} \\ \chit_0^{12} \end{array} \right) \, , \quad
 v^\dagger = \left( \chit_0^{20}, \chit_0^{21} \right) \, .
\end{equation}
To compute the RPA susceptibility
$\chit = \chit_0 \left[ \mathbb{1} - \Gamma_0 \chit_0 \right]^{-1}$
we need to invert
\begin{equation}
 \mathbb{1}_3 - \Gamma_0 \chit_0 = \left( \begin{array}{cc}
 \mathbb{1}_2 - \bar\Gamma_0 \bar\chit_0 & - \bar\Gamma_0 v \\
 - 2U v^\dagger & 1 - 2U \chit_0^{22}
 \end{array} \right)  \, .
\end{equation}
The inverse of a block matrix
\begin{equation}
 M = \left( \begin{array}{cc} A & B \\ C & D \end{array} \right)
\end{equation}
with matrices $A,B,C,D$ can be written as \cite{zhang05}
\begin{equation} \label{schur}
 M^{-1} = \left( \begin{array}{cc}
 A^{-1} + A^{-1} B S^{-1} C A^{-1} & - A^{-1} B S^{-1} \\
 - S^{-1} C A^{-1} & S^{-1} \end{array} \right) \, ,
\end{equation}
where $S = D - C A^{-1} B$ is the so-called Schur complement.
The inverse of $\mathbb{1}_3 - \Gamma_0 \chit_0$ is thus given by Eq.~\eqref{schur} with
$A = \mathbb{1}_2 - \bar\Gamma_0 \bar\chit_0$, $B = - \bar\Gamma_0 v$,
$C = - 2U v^\dagger$, and $D = 1 - 2U \chit_0^{22}$.
Multiplying by $\chit_0$ on the left, one obtains
\begin{equation} \label{eq:chit22q}
 \chit^{22}(\bq,0) =
 v^\dagger(\bq,0) \cdot w(\bq,0) + \chit_0^{22}(\bq,0)/S(\bq,0) \, ,
\end{equation}
where $w = - A^{-1} B S^{-1}$.

$A$ converges to a finite $2 \times 2$ matrix for $\bq \to \bzero$, $B$ and $C$ are linear in $\bq$ for small $\bq$, and $D$ is of order $\bq^2$. Hence, the second term in Eq.~\eqref{eq:chit22q} diverges as $1/\bq^2$ for $\bq \to \bzero$, while the first term tends to a constant and thus becomes irrelevant. Using $\chit_0^{22}(\bzero,0) = (2U)^{-1}$ we thus obtain
\begin{equation}
 \frac{1}{\chit^{22}(\bq,0)} = 2U \left[
 1 - 2U \chit_0^{22}(\bq,0) - 2U v^\dagger(\bq,0) 
 \left[ \mathbb{1}_2 - \bar\Gamma_0 \bar\chit_0(\bzero,0) \right]^{-1} 
 \bar\Gamma_0 \, v(\bq,0) \right] + {\cal O}(|\bq|^3) \, .
\end{equation}
Defining $\bar\Gamma =
\left[ \mathbb{1}_2 - \bar\Gamma_0 \bar\chit_0 \right]^{-1} \bar\Gamma_0$,
one obtains Eq.~\eqref{eq:chit22inv}.


\section{Bare susceptibilities in the N\'eel state}
\label{app:neel}

Since $\bQ$ and $-\bQ$ are equivalent wave vectors in the N\'eel state, the functions $g_\bk$ and $h_\bk$ obey the relations $g_{\bk+\bQ} = g_\bk$ and $h_{\bk+\bQ} = - h_\bk$, respectively, and $e_{\bk+\bQ} = e_{\bk}$.
Hence, the quasi-particle energies $E_\bk^\ell$ and the functions $F_{\ell\ell'}(\bk,\bq,\omega)$ defined in Eq.~\eqref{eq:F} are invariant under a momentum shift by $\bQ$, that is, $E_{\bk+\bQ}^\ell = E_\bk^\ell$ and
$F_{\ell\ell'}(\bk+\bQ,\bq,\omega) = F_{\ell\ell'}(\bk,\bq,\omega)$.

The coherence factors $A_{\ell\ell'}^{02}(\bk,\bq)$, $A_{\ell\ell'}^{03}(\bk,\bq)$, $A_{\ell\ell'}^{12}(\bk,\bq)$, and $A_{\ell\ell'}^{13}(\bk,\bq)$ change sign under a momentum shift $\bk \to \bk+\bQ$. Hence, in the momentum integral in Eq.~\eqref{eq:chit0} for the corresponding bare susceptibilities, contributions from $\bk$ and $\bk + \bQ$ cancel, such that
\begin{equation}
 \chit_0^{02}(\bq,\omega) = \chit_0^{03}(\bq,\omega) =
 \chit_0^{12}(\bq,\omega) = \chit_0^{13}(\bq,\omega) = 0 \, .
\end{equation}
From the obvious relation
$A_{\ell\ell'}^{22}(\bk,\bq) = A_{\ell\ell'}^{33}(\bk,\bq+\bQ)$ one obtains
\begin{equation}
 \chit_0^{22}(\bq,\omega) = \chit_0^{33}(\bq+\bQ,\omega) \, .
\end{equation}
Similarly, $A_{\ell\ell'}^{23}(\bk,\bq) = A_{\ell\ell'}^{23}(\bk,\bq+\bQ)$ yields
\begin{equation}
 \chit_0^{23}(\bq,\omega) = \chit_0^{23}(\bq+\bQ,\omega) \, .
\end{equation}

\end{appendix}



\begin{thebibliography}{99}

\bibitem{goldstone61} J.~Goldstone,
 Field Theories with ``Superconductor'' Solutions,
 Nuovo Cimento {\bf 19}, 154 (1961).

\bibitem{anderson} P.~W.~Anderson,
 {\em Basic Notions of Condensed Matter Physics}
 (Taylor \& Francis, Boca Raton, 1984).
 
\bibitem{halperin69} B.~I.~Halperin and P.~C.~Hohenberg,
 Hydrodynamic Theory of Spin Waves,
 Phys. Rev. {\bf 188}, 898 (1969).
 
\bibitem{halperin77} B.~I.~Halperin and W.~M.~Saslov,
 Hydrodynamic theory of spin waves in spin glasses and other systems with
 noncollinear spin orientations,
 Phys. Rev. B {\bf 16}, 2154 (1977).

\bibitem{harris70} A.~B.~Harris, D.~Kumar, B.~I.~Halperin, and P.~C.~Hohenberg,
 Spin-Wave Damping and Hydrodynamics in the Heisenberg Antiferromagnet,
 J. Appl. Phys. {\bf 41}, 1361 (1970).
 
\bibitem{landau46} L.~D.~Landau,
 On the vibrations of the electronic plasma,
 J. Phys. (USSR), {\bf 10}, 25 (1946).
 
\bibitem{sachdev95} S.~Sachdev, A.~V.~Chubukov, and A.~Sokol,
 Crossover and scaling in a nearly antiferromagnetic Fermi liquid in two dimensions,
 Phys. Rev. B {\bf 51}, 14874 (1995).
 
\bibitem{shraiman89} B.~I.~Shraiman and E.~D.~Siggia,
 Spiral phase of a doped quantum antiferromagnet,
 Phys. Rev. Lett. {\bf 62}, 1564 (1989).

\bibitem{machida89} K.~Machida,
 Magnetism in $\rm La_2CuO_4$ based compounds,
 Physica C {\bf 158}, 192 (1989).

\bibitem{dombre90} T.~Dombre,
 Modulated spiral phases in doped quantum antiferromagnets,
 J. Phys. (France) I {\bf 51}, 847 (1990).

\bibitem{fresard91} R.~Fresard, M.~Dzierzawa, and P.~W\"olfle,
 Slave-Boson Approach to Spiral Magnetic Order in the Hubbard Model,
 Europhys. Lett. {\bf 15}, 325 (1991).

\bibitem{chubukov92} A.~V.~Chubukov and D.~M.~Frenkel,
 Renormalized perturbation theory of magnetic instabilities in the
 two-dimensional Hubbard model at small doping,
 Phys. Rev. B {\bf 46}, 11884 (1992).
 
\bibitem{chubukov95} A.~V.~Chubukov and K.~A.~Musaelian,
 Magnetic phases of the two-dimensional Hubbard model at low doping,
 Phys. Rev. B {\bf 51}, 12605 (1995).

\bibitem{kotov04} V.~N.~Kotov and O.~P.~Sushkov,
 Stability of the spiral phase in the two-dimensional extended $t$-$J$ model,
 Phys. Rev. B {\bf 70}, 195105 (2004).

\bibitem{igoshev10} P.~A.~Igoshev, M.~A.~Timirgazin, 
 A.~A.~Katanin, A.~K.~Arzhnikov, and V.~Yu.~Irkhin,
 Incommensurate magnetic order and phase separation in the two-dimensional Hubbard model
 with nearest- and next-nearest-neighbor hopping,
 Phys. Rev. B {\bf 81}, 094407 (2010).
 
\bibitem{yamase16} H.~Yamase, A.~Eberlein, and W.~Metzner, 
 Coexistence of incommensurate magnetism and superconductivity in the
 two-dimensional Hubbard model, 
 Phys. Rev. Lett. {\bf 116}, 096402 (2016).

\bibitem{eberlein16} A.~Eberlein, W.~Metzner, S.~Sachdev, and H.~Yamase,
 Fermi Surface Reconstruction and Drop in the Hall number due to Spiral
 Antiferromagnetism in High-$T_c$ Cuprates, 
 Phys. Rev. Lett. {\bf 117}, 187001 (2016).	

\bibitem{mitscherling18} J.~Mitscherling and W.~Metzner,
 Longitudinal conductivity and Hall coefficient in two-dimensional metals with spiral
 magnetic order,
 Phys. Rev. B {\bf 98}, 195126 (2018).

\bibitem{bonetti20} P.~M.~Bonetti, J.~Mitscherling, D.~Vilardi, and W.~Metzner,
 Charge carrier drop at the onset of pseudogap behavior in the two-dimensional
 Hubbard model,
 Phys. Rev. B {\bf 101}, 165142 (2020).

\bibitem{qin21} M.~Qin, T.~Sch\"afer, S.~Andergassen, P.~Corboz, and E.~Gull,
 The Hubbard model: A computational perspective,
 arXiv:2104.00064.

\bibitem{sushkov04} O.~P.~Sushkov and V.~N.~Kotov,
 Superconducting spiral phase in the two-dimensional $t$-$J$ model,
 Phys. Rev. B {\bf 70}, 024503 (2004).

\bibitem{inoue95} J.~Inoue and S.~Maekawa,
 Spiral State and Giant Magnetoresistance in Perovskite Mn Oxides,
 Phys. Rev. Lett. {\bf 74}, 3407 (1995).

\bibitem{bao93} W.~Bao, C.~Broholm, S.~A.~Carter, T.~F.~Rosenbaum, G.~Aeppli,
 S.~F.~Trevino, P.~Metcalf, J.~M.~Honig, and J.~Spalek,
 Incommensurate Spin Density Wave in Metallic $\rm V_{2-y}O_3$,
 Phys. Rev. Lett. {\bf 71}, 766 (1993).

\bibitem{takeda72} T.~Takeda, Y.~Yamaguchi, and H.~Watanabe,
 Magnetic structure of $\rm SrFeO_3$,
 J. Phys. Soc. Jpn. {\bf 33}, 967 (1972).

\bibitem{ishiwata20} S.~Ishiwata, T.~Nakajima, J.-H.~Kim, {\em et al.},
 Emergent topological spin structures in the centrosymmetric cubic perovskite
 $\rm SrFeO_3$, Phys. Rev. B {\bf 101}, 134406 (2020).
 
\bibitem{rastelli85} E.~Rastelli, L.~Reatto, and A.~Tassi,
 Quantum fluctuations in helimagnets,
 J. Phys. C {\bf 18}, 353 (1985).
 
\bibitem{chandra90} P.~Chandra, P.~Coleman, and A.~I.~Larkin,
 A quantum fluids approach to frustrated Heisenberg models,
 J. Phys. Condens. Matter {\bf 2}, 7933 (1990).
 
\bibitem{shraiman92} B.~I.~Shraiman and E.~D.~Siggia,
 Excitation spectrum of the spiral state of a doped antiferromagnet,
 Phys. Rev. B {\bf 46}, 8305 (1992).

\bibitem{kampf96} A.~P.~Kampf,
 Collective excitations in itinerant spiral magnets,
 Phys. Rev. B {\bf 53}, 747 (1996).
 
\bibitem{baym61} G.~Baym and L.~P.~Kadanoff,
 Conservation Laws and Correlation Functions,
 Phys. Rev. {\bf 124}, 287 (1961).

\bibitem{negele87} J.~W.~Negele and H.~Orland,
 {\em Quantum Many-Particle Systems} (Addison-Wesley, Reading, MA, 1987).

\bibitem{zhang05} F.~Zhang,
 {\em The Schur Complement and Its Applications} (Springer, New York, 2005).

\end{thebibliography}
\end{document}